\documentclass{aa}
\usepackage{amsmath, amssymb, graphicx, natbib}
\bibpunct{(}{)}{;}{a}{}{,}
\usepackage{txfonts}
\begin{document}

\title{Survival probability and energy modification of hydrogen Energetic Neutral Atoms on their way from the termination shock to Earth orbit}
\author{M. Bzowski } 

\offprints{M.~Bzowski (\email{bzowski@cbk.waw.pl})}

\institute{
Space Research Centre, Polish Academy of Sciences, Bartycka 18A, 00-716 Warsaw, Poland}

\date{}

\abstract
   {Recognizing the transport of Energetic Neutral Atoms (ENA), from their place of birth to Earth orbit, has become an important issue in light of the forthcoming launch of the NASA SMEX mission IBEX, which is devoted to imaging of the heliospheric interface by in-situ detection of ENAs. }
   {We investigate the modifications of both energy of survival probability of the hydrogen ENA (H ENA) detectable by IBEX (0.01 -- 6~keV), between the termination shock and Earth orbit. We take into account the influence of the variable and anisotropic solar wind and of solar EUV radiation. 
}
   {Energy changes of the atoms are calculated by numerical simulations of the orbits of H ENA between $\sim 100$~AU from the Sun and Earth orbit, taking into account solar gravity and Lyman-$\alpha$ radiation pressure, which is variable in time and depends on the radial velocity of the atom. To calculate the survival probabilities of the atoms against ionization, a detailed observation-based 3D and time-dependent model of H ENA ionization is constructed, and with the use of this model the probabilities of survival of the atoms are calculated by numerical integration along the previously-calculated orbits.}
   {Due to radiation pressure, H ENA reach the Earth orbit practically without energy and direction change, apart from the atoms of energy lower than 0.1~keV, during high solar activity. The survival probability of H ENA increases from just $\sim 2\%$ for the slowest detectable ENA at solar minimum to $\sim 80$\% for the fastest ENA. For a given energy at Earth orbit we expect fluctuations in the survival probability of amplitude between $\sim 20$ percent at 0.01~keV to just a few percent at 6~keV and a modulation of survival probability as a function of the location at Earth orbit, ecliptic latitude of the arrival direction, and phase of solar cycle with an amplitude of a few dozen percent for 0.1~keV atoms at solar minimum to a few percent for 6~keV atoms at solar maximum. 
}
   {With appropriate account of local transport effects IBEX should be able to discover departures from symmetry in the flux of H ENA from the heliospheric interface at a level of a few percent.}


\titlerunning{Modifications of H ENA between termination shock and Earth orbit }
\authorrunning{M.Bzowski}

\maketitle

\section{Introduction}
Hydrogen Energetic Neutral Atoms (H ENA) provide important information about the physical state of plasma in remote parts of the heliosphere, especially in the heliospheric interface \citep{gruntman:92b, hsieh_etal:92a, gruntman:97, gruntman_etal:01a, scherer_fahr:03a, fahr_scherer:04a, heerikhuisen_etal:07a, fahr_etal:07a, sternal_etal:08a}. Measurements of heliospheric H ENA are planned by the NASA SMEX mission IBEX \citep{mccomas_etal:04a, mccomas_etal:05a, mccomas_etal:06}. IBEX will be a spin-stabilized Earth satellite observing Energetic Neutral Atoms (ENA) in the 0.01 -- 6~keV energy band, with sensors looking perpendicularly to the spin axis directed at the Sun. The prime target of the mission will be imaging of the heliospheric interface via neutral atoms and the goal of an individual observation will be to obtain the kinematic parameters of the locally-registered atom beyond the termination shock (TS). To that end, understanding of the processes affecting the atoms between their source region and the detector is needed. In particular, a control on the modifications of the kinematic parameters of the atoms between the source region and the detector and a grasp on loss processes along the way are necessary. Some insights into this problem were presented by \citet{bzowski_tarnopolski:06a}. 

Since the ENA gas inside the TS is collisionless at all energies, then -- from Boltzmann equation -- the transport of ENA from TS to the inner heliosphere is governed solely by dynamic effects and gain and loss processes. The dynamic effects include heliocentric acceleration by a combined action of the solar gravity and the opposing solar radiation pressure, whose strength depends on the radial velocity of the incoming atom. The losses include all known ionization processes operating in the heliosphere: charge exchange, EUV ionization, and electron impact \citep{rucinski_etal:96a}. For inwardly-traveling ENA, the gains (considered by IBEX as unwanted foreground) include charge exchange between CME, CIR, and pickup ions on the one hand, and all kinds of neutral populations present inside TS on the other hand, as well as production of ENA in the plasma environment of the spacecraft. In this paper, they are neglected because they will constitute a small addition to the net signal \citep{fahr_etal:07a}.

The paper discusses modifications of kinematic parameters and survival probabilities of H ENA, suitable for detection by IBEX, traveling from the heliospheric termination shock (TS) to 1~AU. First, we discuss the radiation-pressure force affecting the atoms and the resulting trajectories of individual H ENA reaching perihelia of their trajectories at Earth orbit. Then we briefly present the ionization processes destroying the incoming H ENA will be briefly presented, providing a longer discussion in the Appendix. Subsequently, we show the survival probabilities of the H ENA, with emphasis on asymmetries and anisotropies that result from local conditions in the solar wind. The discussion is based, as much as possible, on available solar-wind and solar-radiation output data and does not consider measurement technicalities, such as detection efficiencies, geometric factors, proper motion of the spacecraft etc., which are left to a future study. Emphasis is placed on atoms at the energies expected in the heliosheath region, i.e. up to 0.2~keV; faster atoms are not, however, disregarded, since the studies mentioned earlier predict a non-vanishing production  of H ENA to the upper limit of the IBEX sensitivity threshold. Attention is drawn to solar-cycle-related effects, which include changes in the solar-wind anisotropy and net flux of solar Lyman-$\alpha$ pressure, and to fluctuations on a timescale of the solar rotation period.

The term ``source region'' frequently used in the paper should be understood as an origin of the H ENA, sufficiently distant from the Sun for its dynamical influence on the atoms to be negligible. It was arbitrarily set to be $\sim 100$~AU from the Sun. An atom that is unaffected by any radiation pressure and at infinity travels at 12~km/s, at 1~AU will have accelerated to 43.8~km/s, while an atom that travels with the velocity 12~km/s at 100~AU from the Sun, at 1~AU will have accelerated to 43.6~km/s  and the difference between the velocities is then just 0.2~km/s; when radiation pressure is added and larger velocities are adopted, this difference will decrease further. 

Dynamical conclusions discussed in Sect.~2 are valid for all atoms reaching perihelion at 1~AU from the Sun irrespective of exact location in the 3D space; the conclusions about the survival probability, however, presented in Sect.~3 were obtained for H ENA reaching perihelia of their trajectories at 1~AU in the ecliptic plane, which is referred to as ``at Earth orbit''. 

\begin{figure*}
\centering
\includegraphics[width=8cm]{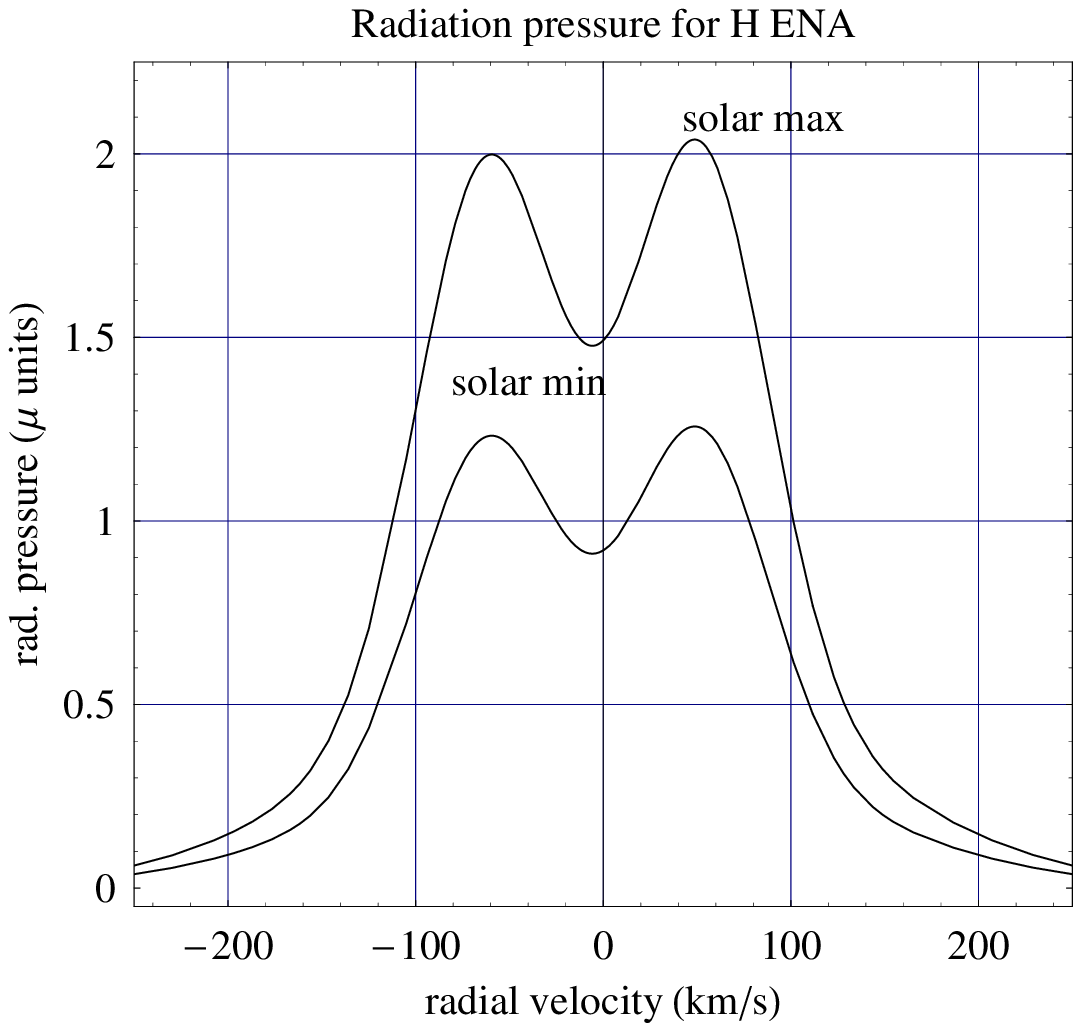}
\includegraphics[width=8cm]{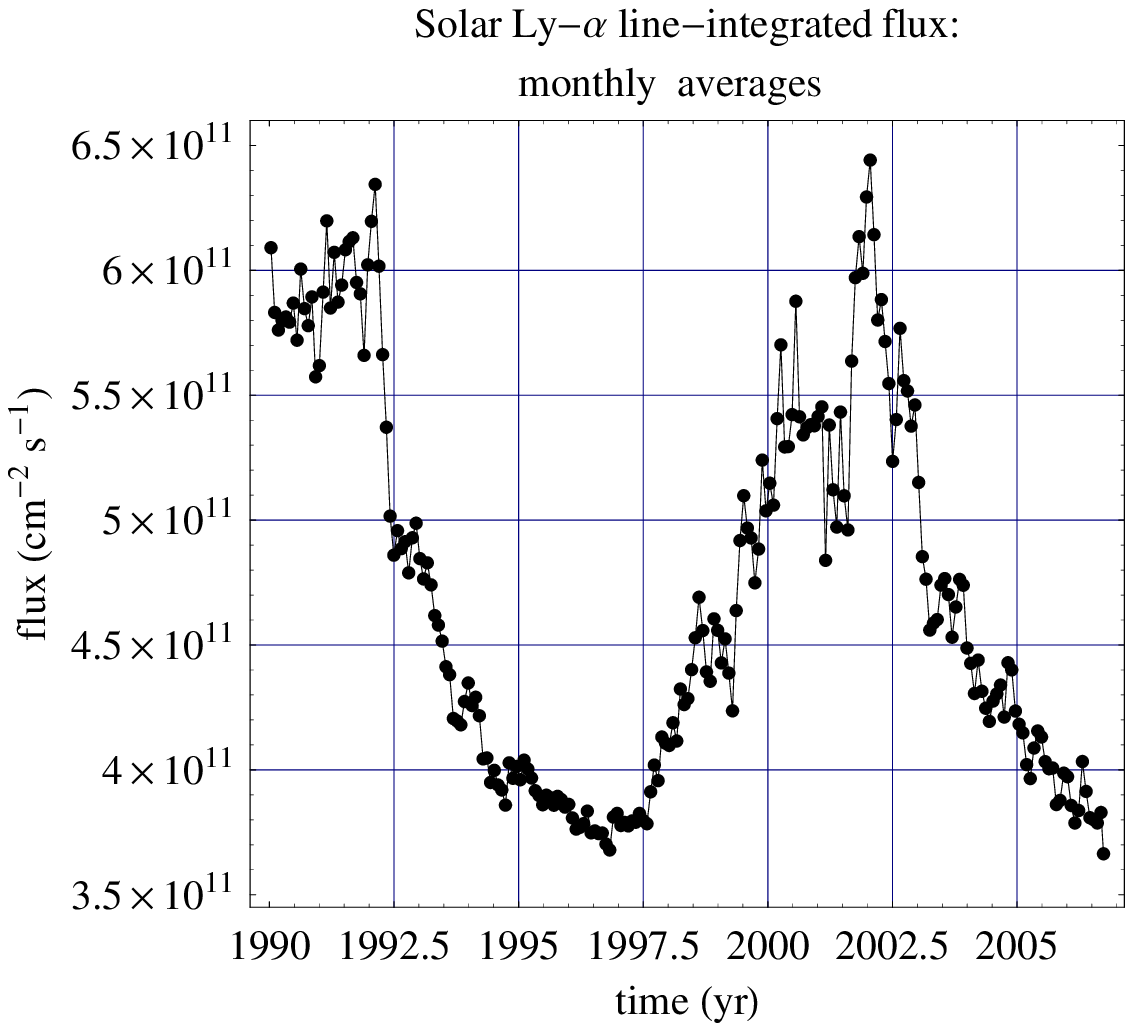}
\caption{Radiation pressure acting on H atoms, expressed as a factor $\mu$ of compensation of the solar gravity force, shown as a function of their radial velocity with respect to the Sun for solar minimum and maximum (left-hand panel), and wavelength- and disk-integrated monthly-averaged solar Lyman-$\alpha$ flux based on the SOLAR 2000 model (right-hand panel). The profiles of radiation-pressure factor $\mu$ are shown according to the model defined in Eq.(\ref{e3}) for the net flux $I_{\mathrm{min}} = 3.7 \cdot 10^{11}$ cm$^{-2}$~s$^{-1}$ (solar min.) and $I_{\mathrm{max}} = 6.0 \cdot 10^{11}$ cm$^{-2}$~s$^{-1}$ (solar max.).}
\label{lyaPlot}
\end{figure*}
\begin{figure*}
\centering
\includegraphics[width=8cm]{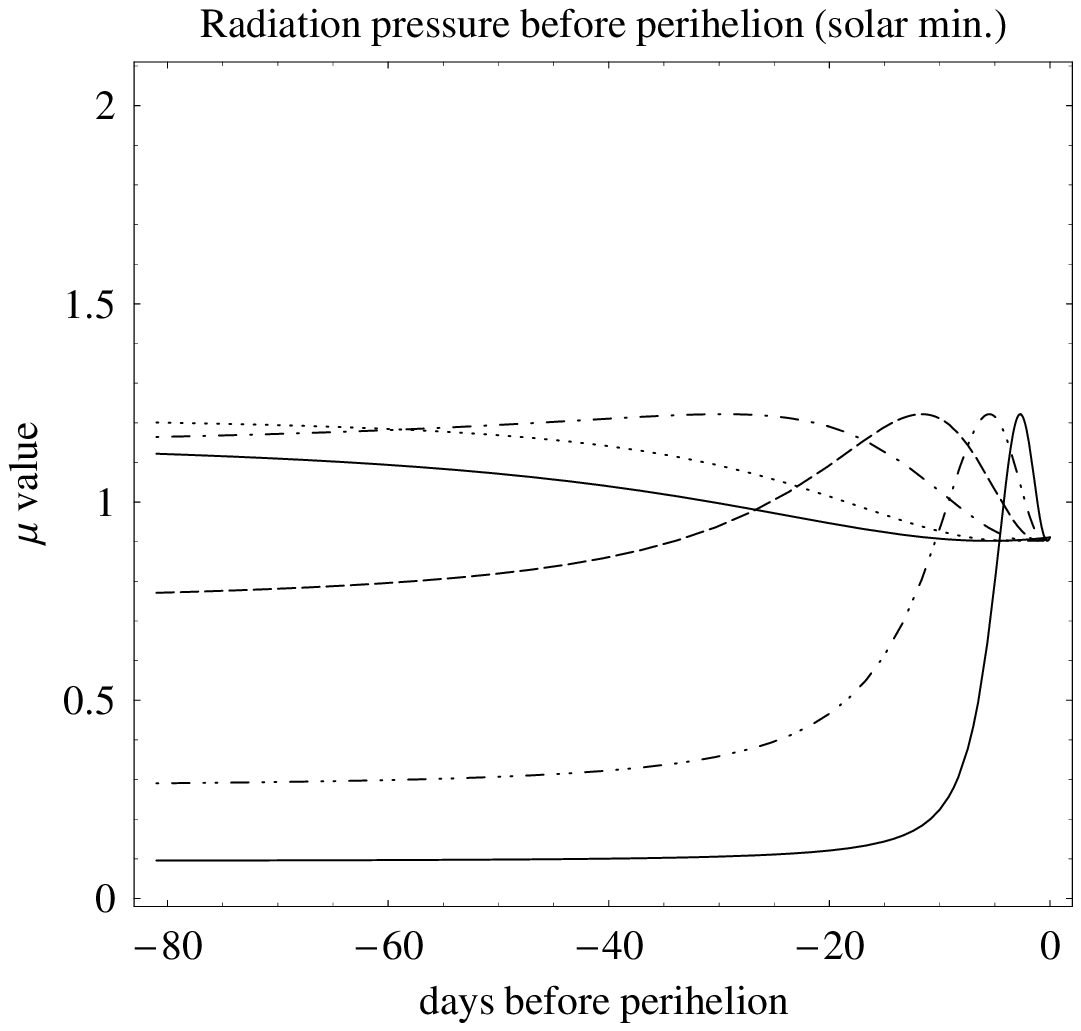}
\includegraphics[width=8cm]{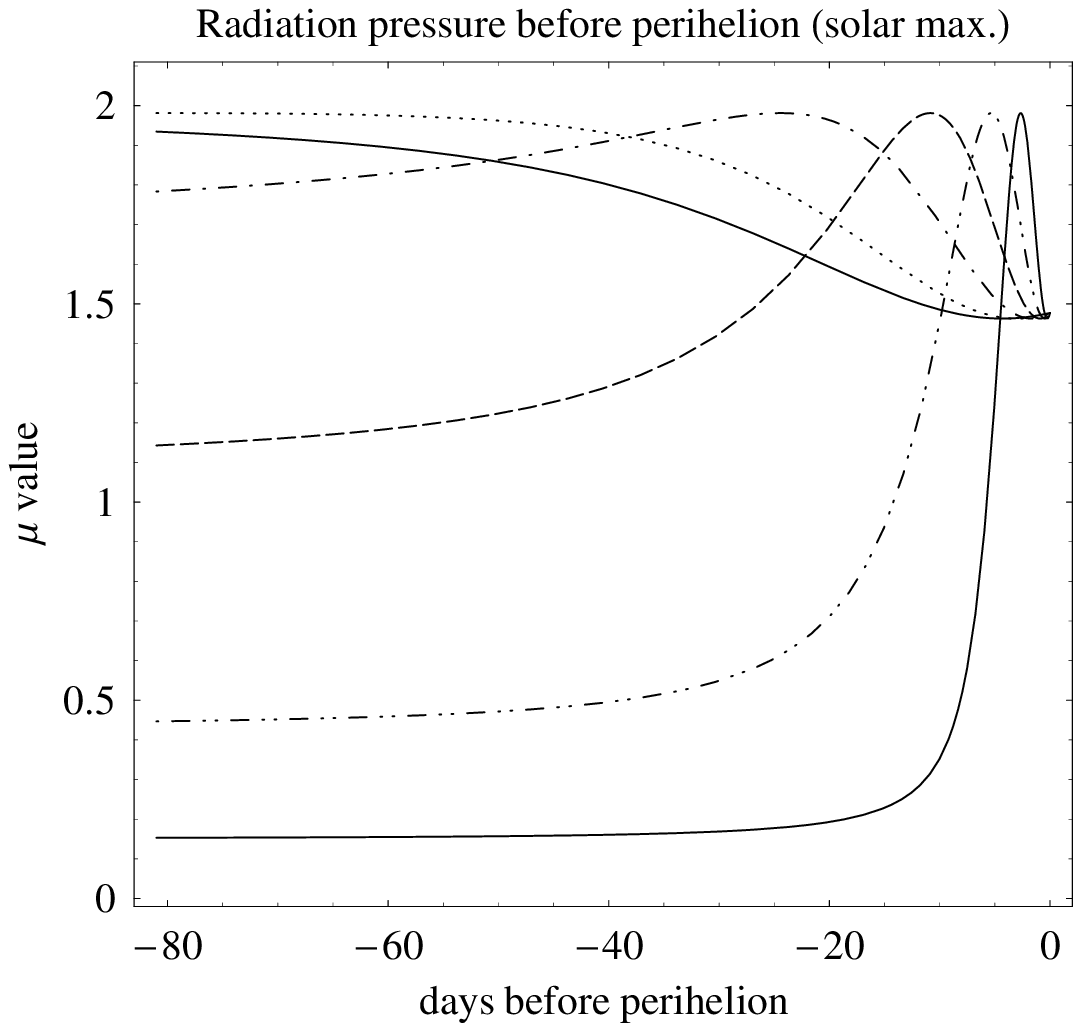}
\caption{Instantaneous radiation pressure acting on H ENA during 81 days before perihelion located at 1~AU, expressed as the gravity-compensation factor $\mu$ and shown for a few energies corresponding to the lowermost IBEX energy channels and equal to 0.01~keV (solid), 0.015~keV (dots), 0.029~keV (dash-dot), 0.056~keV (dashed), 0.107~keV (dash-dot-dot), 0.208~keV (lowermost solid lines), for solar minimum (left-hand panel), and maximum conditions (right-hand panel). The perihelion velocities correspond, respectively, to 43.8, 53.6, 74.5, 103.6, 143.2, and 199.6~km/s.  }
\label{radPresPlot}
\end{figure*}
\section{Dynamics of H ENA atoms between the termination shock and inner heliosphere}
\subsection{Radiation pressure and equation of motion}
The velocities of heliospheric H ENA that can be observed by IBEX are in the approximate range $\sim 50 - 1000$~km/s, which corresponds to an energy range 0.01 -- 6~keV \citep{mccomas_etal:04a, mccomas_etal:05a, mccomas_etal:06}. The motion of such atoms between the termination shock and inner heliosphere is determined on the one hand by the solar gravity force, and on the other hand by resonance interaction with the solar Lyman-$\alpha$ radiation. The latter phenomenon produces a force called radiation pressure, which decreases with heliocentric distance following $1/r^2$, similar to solar gravity and has a radial direction that is away from the Sun. Hence, it is convenient to express the radiation pressure as a factor $\mu$ of (over)compensation of solar gravity. The solar Lyman-$\alpha$ line is about 0.15~nm wide and features a self-reversed profile with a central trough \citep{lemaire_etal:02, lemaire_etal:05}. Given the non-flat shape of the solar Lyman-$\alpha$ line (see Fig.~\ref{lyaPlot}) on one hand, and the range of radial velocities of the incoming H ENA atoms on the other hand one expects that, due to the Doppler effect, the radiation pressure acting on the atoms will be a non-negligible function of their radial velocities with respect to the Sun. If the heliospheric gas is optically thin, all neutral atoms in the heliosphere obey the equation of motion in the form:
\begin{equation}
\label{equ2}
d^2 \vec{r}/{dt^2} = -G\, M \left[1 - \mu\left(v_r\left(\vec{r}\right), I_{\mathrm{ tot}}\left(t, \phi\right)\right)\right] \vec{r}/|\vec{r}|^3
\end{equation}
where $\vec{r}\left(t\right)$ is the position vector of the atom with respect to the Sun at a time $t$, $v_r = \left(\mathrm{d} \vec{r}/\mathrm{d}t\right) \cdot \vec{r}/|\vec{r}|$ is its radial velocity, and $G\,M$ is the product of the gravity constant and the solar mass. $I_{\mathrm{ tot}}\left(t, \phi\right)$ is the wavelength-integrated solar Lyman-$\alpha$ flux at a time $t$ and heliolatitude $\phi$, and $\mu$ is the radiation pressure expressed as a compensation factor of solar gravity. According to \citet{scherer_fahr:96} and \citet{quemerais:00}, the optical depth of the neutral interstellar gas inside the heliosphere cannot be neglected beyond $\sim 10$~AU from the Sun in the case of atoms of thermal energies ($<\sim 0.01$~keV); for radial velocities exceeding $\sim 30$~km/s, however,  the gas is optically thin and the above-mentioned approximation is valid. 

Based on a fit of Eq.~(\ref{e3}) to 9 solar Lyman-$\alpha $ line profiles observed by \citet{lemaire_etal:02}, the solar Lyman-$\alpha$ line profile can be parameterized by a simple function of wavelength- and disk-integrated flux $I_{\mathrm{tot}}$ (Tarnopolski, 2008, thesis;  \cite{tarnopolski_bzowski:07a, bzowski_etal:08a}):
\begin{eqnarray}
\label{e3}
\mu \left(v_r, I _{\mathrm{tot}}\right)&=&A\left(1 + B\, I_{\mathrm{tot}}\right) \exp \left(-C v_r^2\right)\\ &\times &\left[1 + D \exp \left(-F v_r
- G v_r^2\right) + H \exp \left(P v_r - Q v_r^2\right)\right] \nonumber
\end{eqnarray}
where $A, B, C, D, F, G, H, P$, and $Q$ are parameters of the fit, which are compiled in Table \ref{profTable}. Conclusions presented in the following part of the paper are based on numerical solutions of the equation of motion priovided in Eq.~(\ref{equ2}) in connection with Eq.~(\ref{e3}), which were obtained with initial conditions including velocity vectors perpendicular to the local radial direction at a starting point located at Earth orbit, with velocity magnitudes corresponding to energy channels planned for the IBEX detectors.
\begin{table}
\caption{Parameters of the model of radiation-pressure dependence on radial velocity $v_r$, expressed in km/s, and on total flux $I_{\mathrm{tot}}$, expressed in cm$^{-2}$~s$^{-1}$, defined in Eq.(\ref{e3}).}
\label{profTable}
\centering
\begin{tabular}{lllllllll}
\hline
 $A  =  2.4543 \cdot 10^{-9}$, & $B =  4.5694\cdot 10^{-4}$, & $C =  3.8312\cdot 10^{-5}$, \\ $D =  0.73879$, & $F =  4.0396\cdot 10^{-2}$, &  $G =  3.5135\cdot 10^{-4}$, \\ 
 $H =  0.47817$, &  $P =  4.6841\cdot 10^{-2}$, &  $Q =  3.3373\cdot 10^{-4}$\\
\hline
\end{tabular}
\end{table}

Observations of $I_{\mathrm{tot}}$ have been performed for about three solar cycles and inevitable gaps were filled with proxy values \citep{tobiska_etal:00c}. The absolute calibration is still somewhat uncertain. In present paper, as basis for $I_{\mathrm{tot}}$ the SOLAR 2000 model was used \citep{woods_etal:00}, which was also employed by \citet{lemaire_etal:02} to calibrate their observed solar Lyman-$\alpha$ line profiles. In the calculations the Lyman-$\alpha$ flux was assumed to be constant in time and for solar minimum to be equal to $I_{\mathrm{min}} = 3.7 \times 10^{11}$ cm$^{-2}$~s$^{-1}$ and for solar maximum to $I_{\mathrm{max}} = 6.0 \cdot 10^{11}$ cm$^{-2}$~s$^{-1}$.
\begin{figure*}
\centering
\includegraphics[width=8cm]{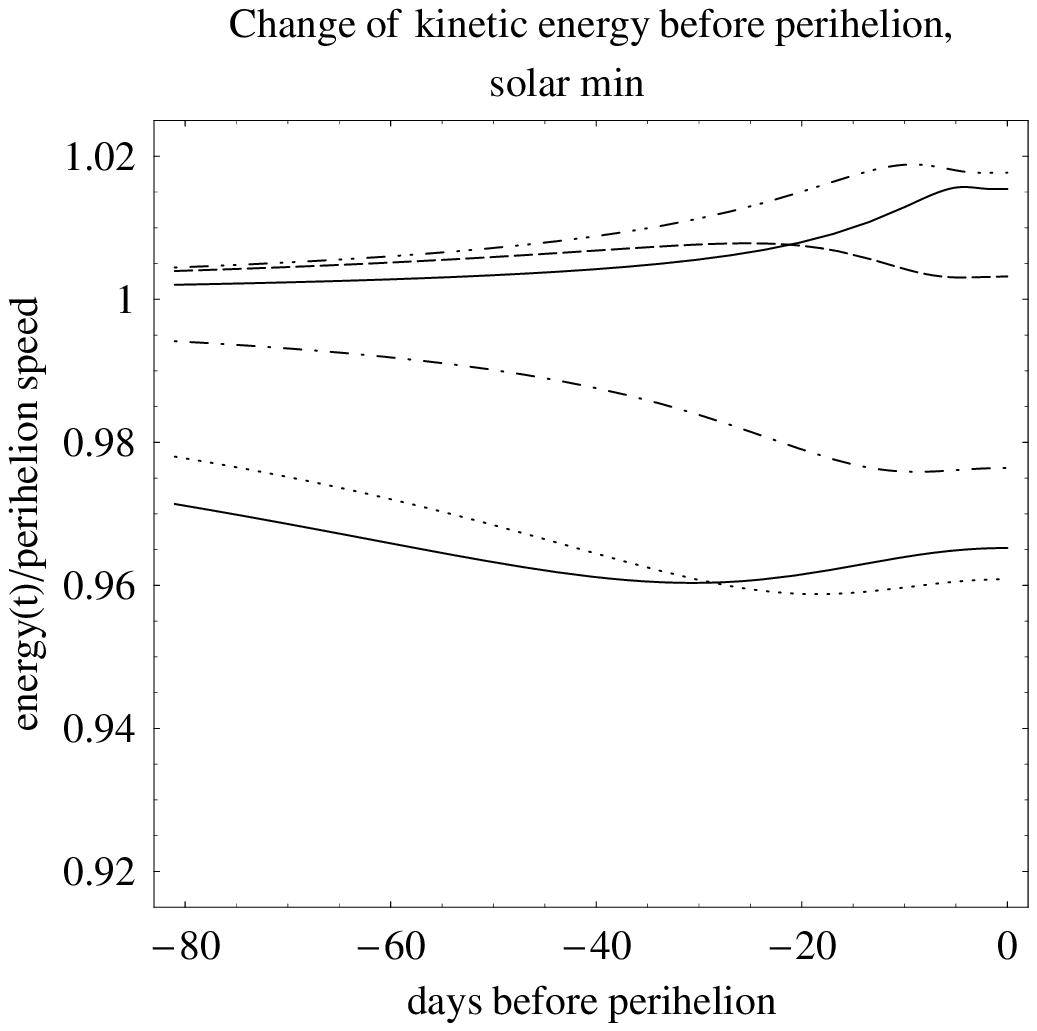}	
\includegraphics[width=8cm]{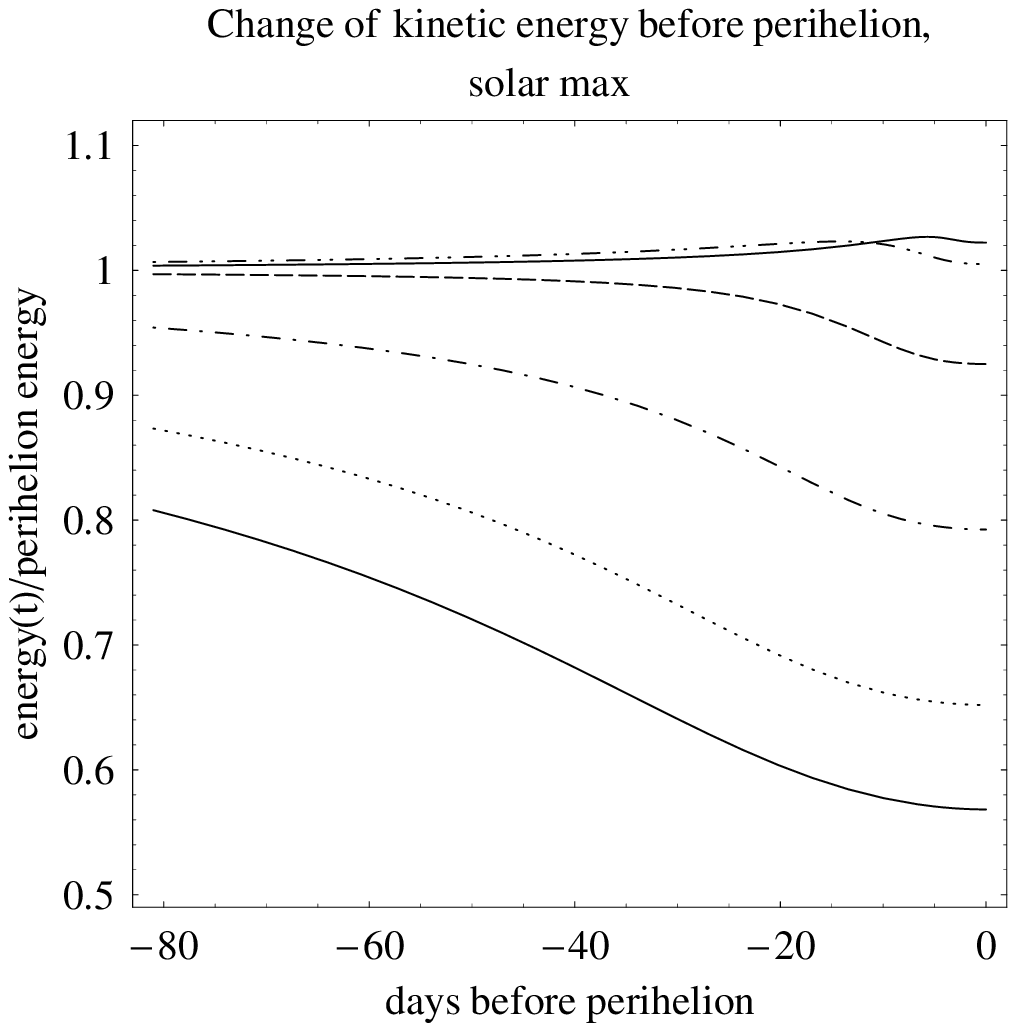}
\caption{Ratios of energy of H ENA at time $t$ before perihelion located at 1~AU to the energy in the source region during 81 days before perihelion for a few 1~AU energies, corresponding to the lowermost IBEX energy channels and equal to 0.01~keV (solid), 0.015~keV (dots), 0.029~keV (dash-dot), 0.056~keV (dashed), 0.107~keV (dash-dot-dot), 0.208~keV (solid lines), for solar minimum (left-hand panel) and maximum conditions (right-hand panel).  }
\label{EnChangePl}
\end{figure*}

\subsection{Trajectories of H ENA and modifications of their energies between the source region and 1~AU}
\begin{figure}
\includegraphics[width=8cm]{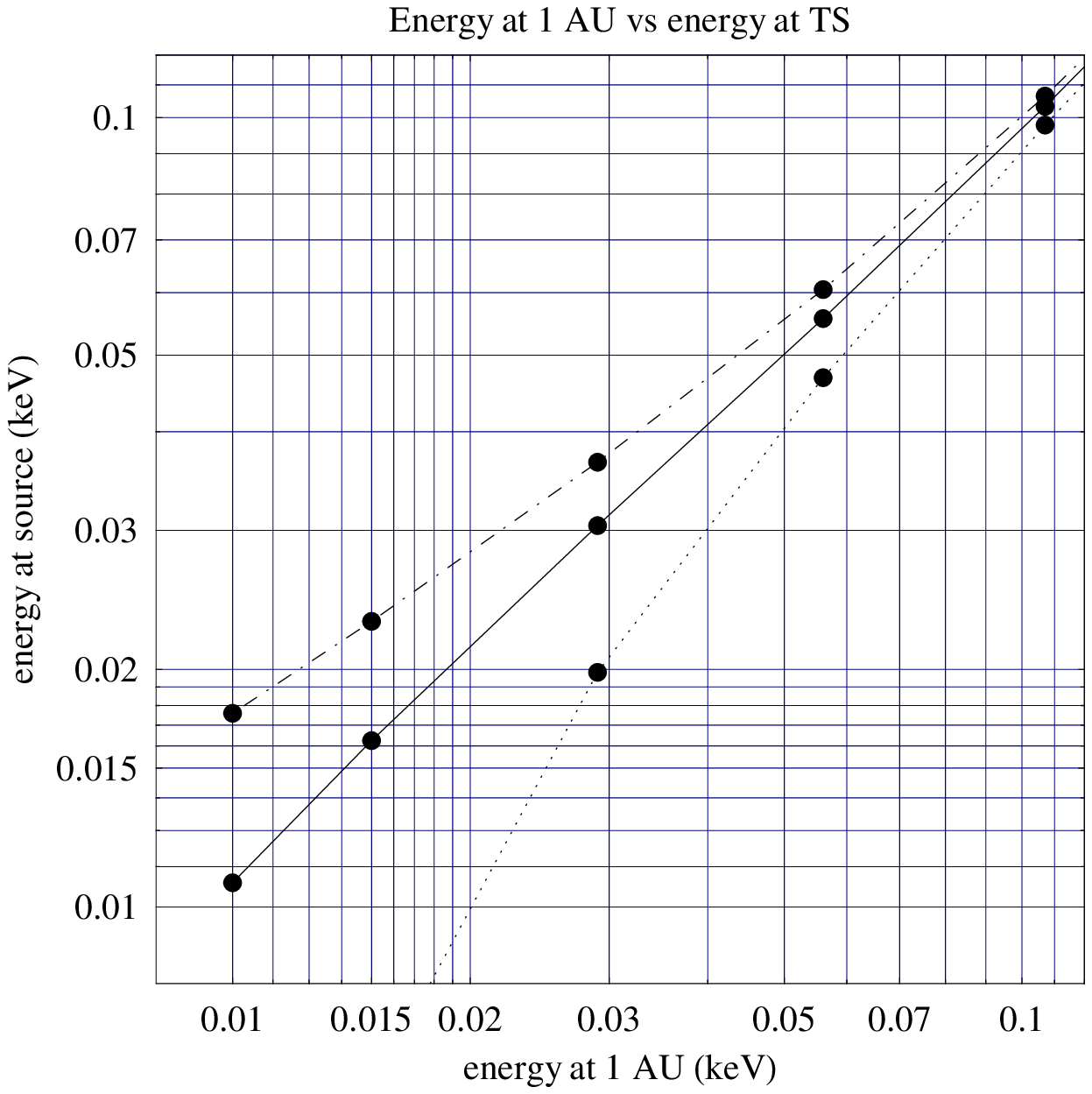}
\caption{Energy of H ENA at 1~AU versus energy in the source region at 100~AU. Solid line: solar-minimum conditions, broken line: solar maximum-conditions, dotted line: purely ballistic case with no radiation pressure, shown for comparison. Projections of dots on the horizontal axis mark the IBEX energy channels, apart from the lowermost one, which marks the low-energy sensitivity limit. }
\label{sourceEnClipped}
\end{figure}

\begin{figure*}
\centering
\includegraphics[width=8cm]{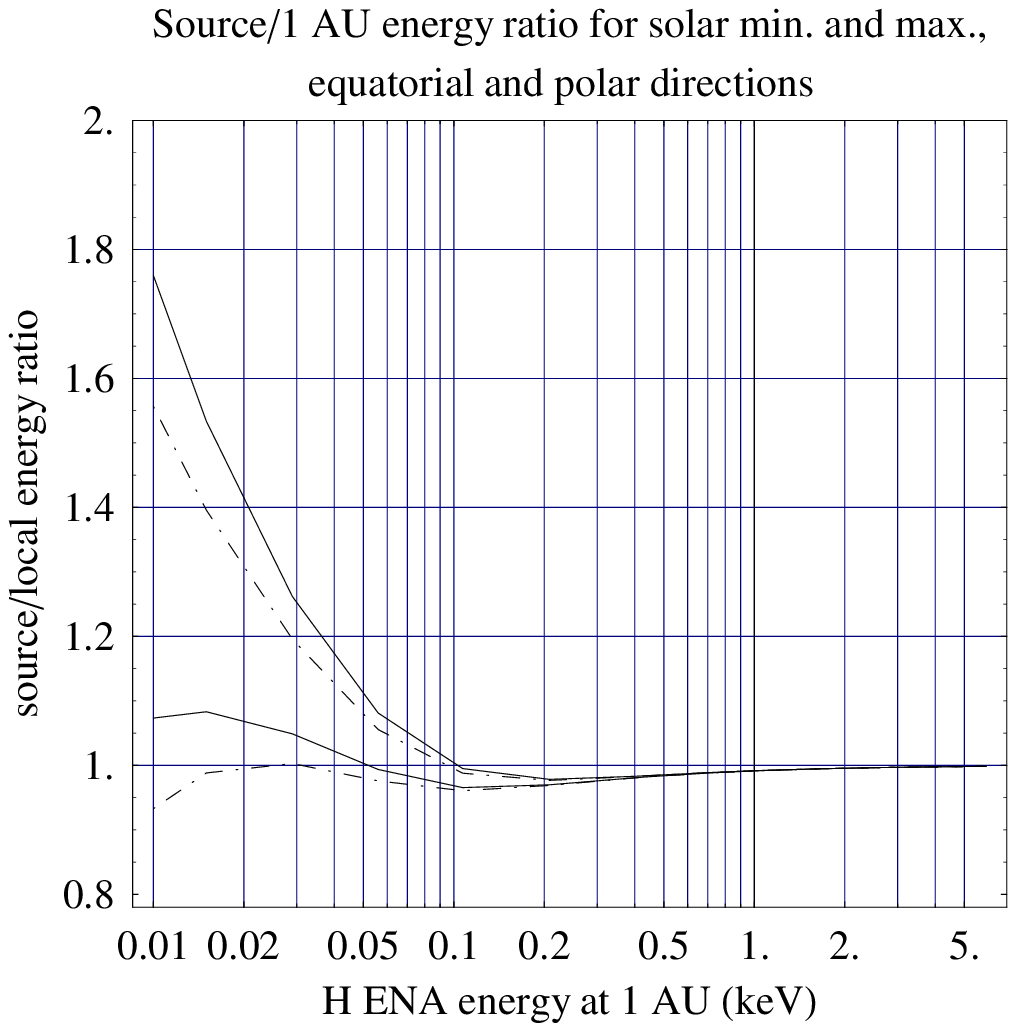}
\includegraphics[width=8cm]{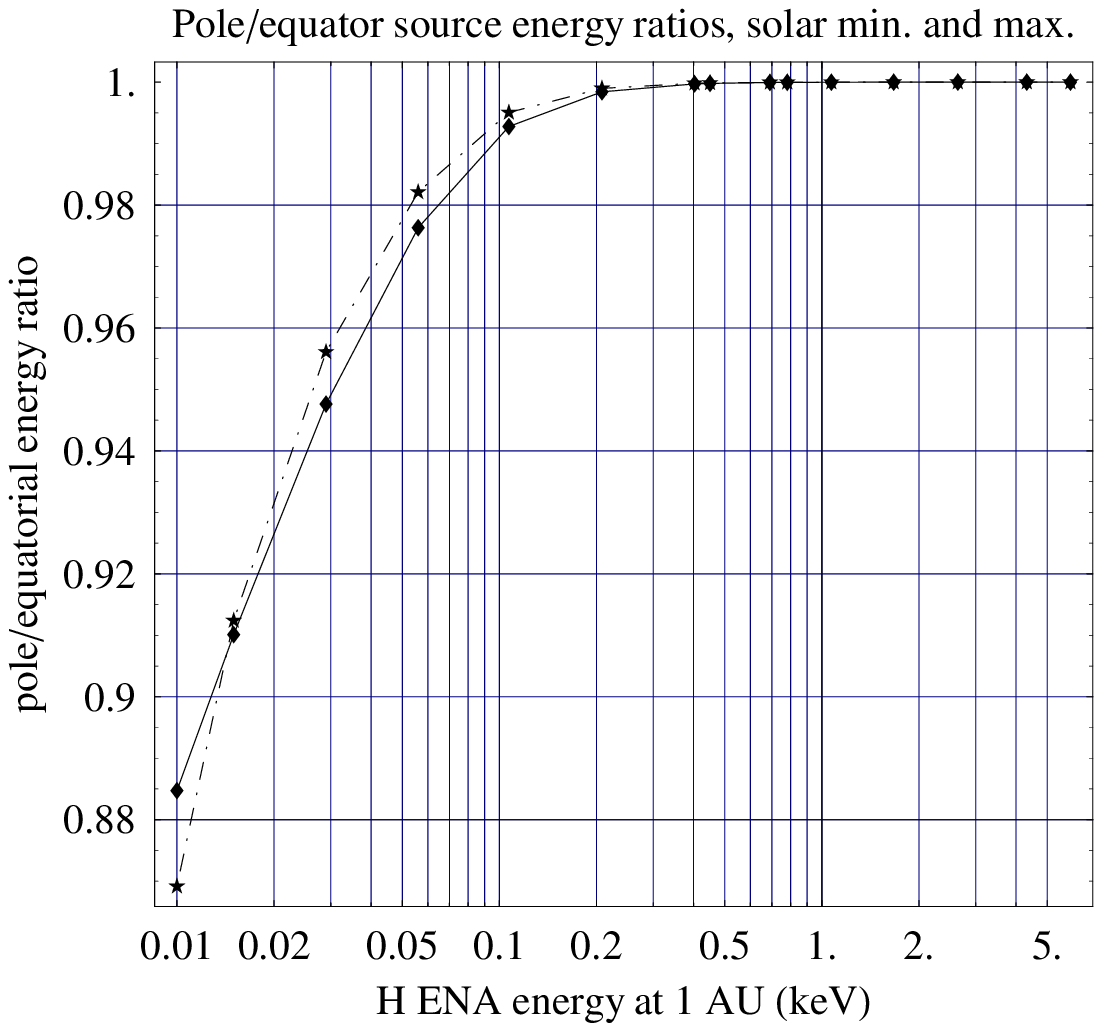}
\caption{{\em Left-hand panel}: Ratios of H ENA energy in the source region $E_{\mathrm{S, eq}}$ to their energy at 1~AU for atoms incoming from polar regions $E_{\mathrm{S}}/E_{\mathrm{1 AU}}$ (broken lines) and from equatorial regions $E_{\mathrm{eq}}/E_{\mathrm{1 AU}}$ (solid lines) during solar minimum and maximum conditions. The upper pair of lines corresponds to solar maximum, the lower pair to solar-minimum conditions. {\em Right-hand panel}: ratios of energies in the source region of atoms incoming from the poles and from an equatorial band $E_{\mathrm{S}}/E_{\mathrm{eq}}$, shown as a function of their energy at 1~AU for solar minimum and maximum conditions (respectively, solid and broken line). } 
\label{enRatiosPl}
\end{figure*}

\begin{figure}
\centering
\includegraphics[width=8cm]{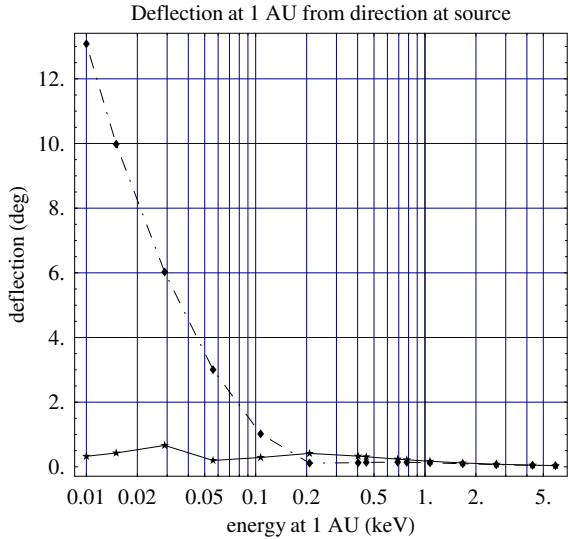}
\caption{Difference between the original direction of ENA within the source region and the direction at 1~AU for solar minimum conditions (solid line) and solar maximum (broken line) as a function of energy at 1~AU.}
\label{deflectionPlot}
\end{figure}

Because of the geometry of its lines of sight, IBEX will measure only the inward-traveling atoms that reach perihelion at 1~AU. Such atoms are detected at approximately zero radial velocity with respect to the Sun, but for most of their paths they have radial velocities that are almost equal to their net velocities. If their net velocity in the source region is larger than the spectral range of the solar Lyman-$\alpha$ line, the atoms do not experience any radiation pressure underway, aprat from within a few AU from the Sun, where, before reaching perihelion, they their radial velocities will be reduced to zero (see Fig.~\ref{radPresPlot}). Since the velocities are large, however, the trajectories will be weakly modified by solar gravity and remain close to straight lines. On the other hand, slower ENA (whose velocities remain within the spectral range of the solar line) will be stronger affected by radiation pressure. Effectively, almost all changes in the kinetic energy of incoming atoms occur within $\sim 3$ solar rotations before the 1~AU perihelion or an even shorter time (the faster the atom moves, the shorter is the effective interaction time), as illustrated in Fig.~\ref{EnChangePl}. During solar minimum, the effective radiation pressure is, however, close to $\mu = 1$, which produces little modification of the trajectories and energies. In contrast, the trajectories of the slowest atoms in the discussed energy range will be affected considerably during solar maximum, when radiation pressure overcompensates significantly for solar gravity. Modification of their energies between the source region and perihelion at 1~AU is shown in Fig.~\ref{sourceEnClipped} and in the left-hand panel of Fig.~\ref{enRatiosPl}. 
 The figures demonstrate that energy modifications are negligible for all IBEX energy intervals apart from the slowest atoms during solar-maximum conditions: in this case, they can reach 50\%. Atoms faster than ~200 km/s at 1~AU  practically do not change their energy. The difference between the arrival direction and the direction at the source region is negligible at solar minimum (see Fig.~\ref{deflectionPlot}), but at the lowest energies increases to 14\degr\, at solar maximum. Deflections larger than the $7\degr$ of the IBEX pixel width are therefore possible only in the case of energies below $\sim 0.03$~keV.

As is evident from Fig.~\ref{lyaPlot}, the solar Lyman-$\alpha$ flux exhibits appreciable fluctuations on both monthly and daily scales. Analysis of daily time series, split into Carrington-rotation intervals, demonstrates that the ratios of standard deviations of the Carrington-averaged values to the Carrington-averaged values vary from $\sim 2$\% at solar minimum to $\sim 5$\% at solar maximum. Based on the results presented above, it is surmised that such fluctuations should not introduce appreciable fluctuations in the energies and directions of the incoming ENA atoms. 

\subsection{Effects of latitudinal anisotropy of the Lyman-$\alpha$ flux} 

Potentially stronger variation than the fluctuations discussed in the previous section might be expected because of the latitudinal anisotropy in the solar Lyman-$\alpha$ output. As proposed by \citet{cook_etal:80a,cook_etal:81a} and observed directly by \citet{auchere:05} during solar minimum and indirectly by \citet{pryor_etal:92} during solar maximum, the line-  and disk-integrated solar Lyman-$\alpha$ flux is a function of heliolatitude. The theory and limited observations available appear to imply that the polar flux $I_{\rm pol}$ should be equal to about 0.8 of the equatorial flux $I_{\rm eqtr}$ and that this ratio should not change appreciably during the solar cycle. More research is certainly required to establish details of this phenomenon and the anisotropy-related conclusions presented below must be treated with caution. By adopting this anisotropy and assuming that the symmetry plane of the latitudinal pattern of the Lyman-$\alpha$ flux is the solar equator and the line profile does not change with latitude, we would expect a yearly modulation of the flux of H ENA originating in polar regions (because of the Earth travel in heliolatitude by $\pm 7.25\degr$) and a differentiation of H ENA energies as a function of arrival-direction helioaltitude.

To simulate the ``flattening'' of the net Lyman-$\alpha$ flux at polar latitudes, the radiation-pressure compensation-factor $\mu$ in Eq.(\ref{e3}) was modified to be 
\begin{equation}
\label{e4}
 \mu\left(\phi\right) = \mu\left(0\right)\sqrt{\cos^2 \phi + f^2 \sin^2 \phi}
\end{equation}
 Simulations demonstrated that for a selected H ENA energy at 1~AU (denoted $E_{1\, \mathrm {AU}}$), the modulation of the ratio of H ENA energies, in the source region at the ecliptic poles $E_{\mathrm N}/E_{\mathrm S}$, as a function of ecliptic longitude of perihelion for the lowest energies detectable by IBEX $E_{\mathrm{1\,AU}} = 0.01$~keV, is approximately 4\%, and at 0.06~keV, is already as small as 1\%. This effect can therefore be considered to be negligible. 

Irrespective of the solar-cycle phase, the source-region energies $E_{\mathrm{ src}}\left(\phi\right)$ of the H ENA arriving at 1~AU from different latitudes $\phi$ with the same energy $E_{\mathrm{1\,AU}}$, vary with $\phi$ by no more than $\sim 10$\% in the case of lowest energies (see the right-hand panel of Fig.~\ref{enRatiosPl}). These variations decrease to below 5\% at $E_{\mathrm{1\,AU}} \simeq 0.02$~keV and disappear altogether for $E_{\mathrm{1\,AU}} > 0.2$~keV.

The effects of anisotropy in the solar Lyman-$\alpha$ flux persist both during solar minimum and solar maximum. During solar minimum, the atoms originating in the polar regions are transmitted with a very small energy change for the entire energy range (see the broken line in the solar-minimum pair of lines in the left-hand panel of Fig.~\ref{enRatiosPl}), while a 20\% change in the energy of the slowest atoms originating in ecliptic latitudes is expected (the solid line). Similarly during solar maximum, the energy change for polar directions is smaller than for equatorial directions (see the upper pair of lines in the left-hand panel of Fig.~\ref{enRatiosPl}). 

\begin{figure*}
\centering
\includegraphics[width=8cm]{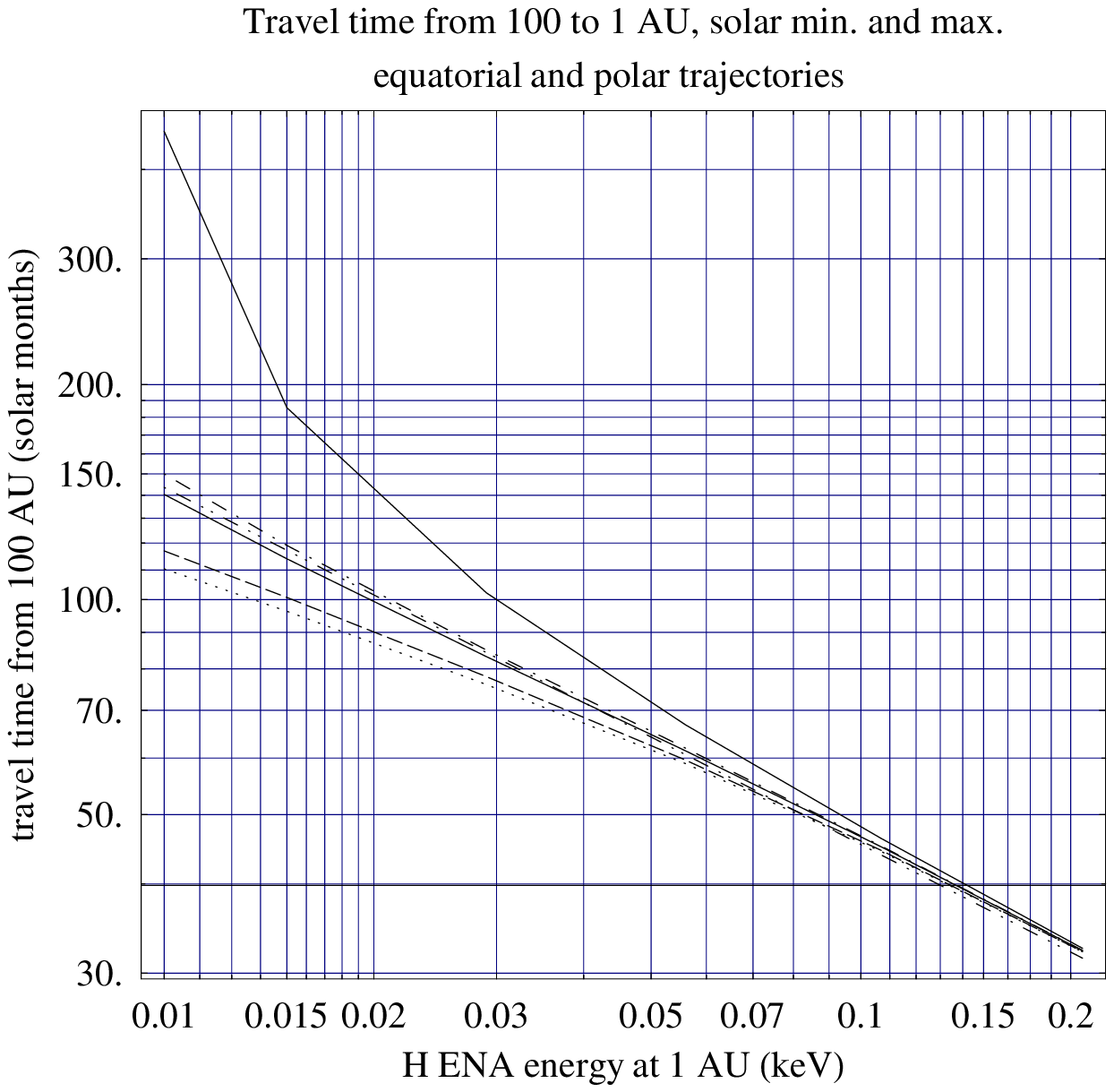}
\includegraphics[width=8cm]{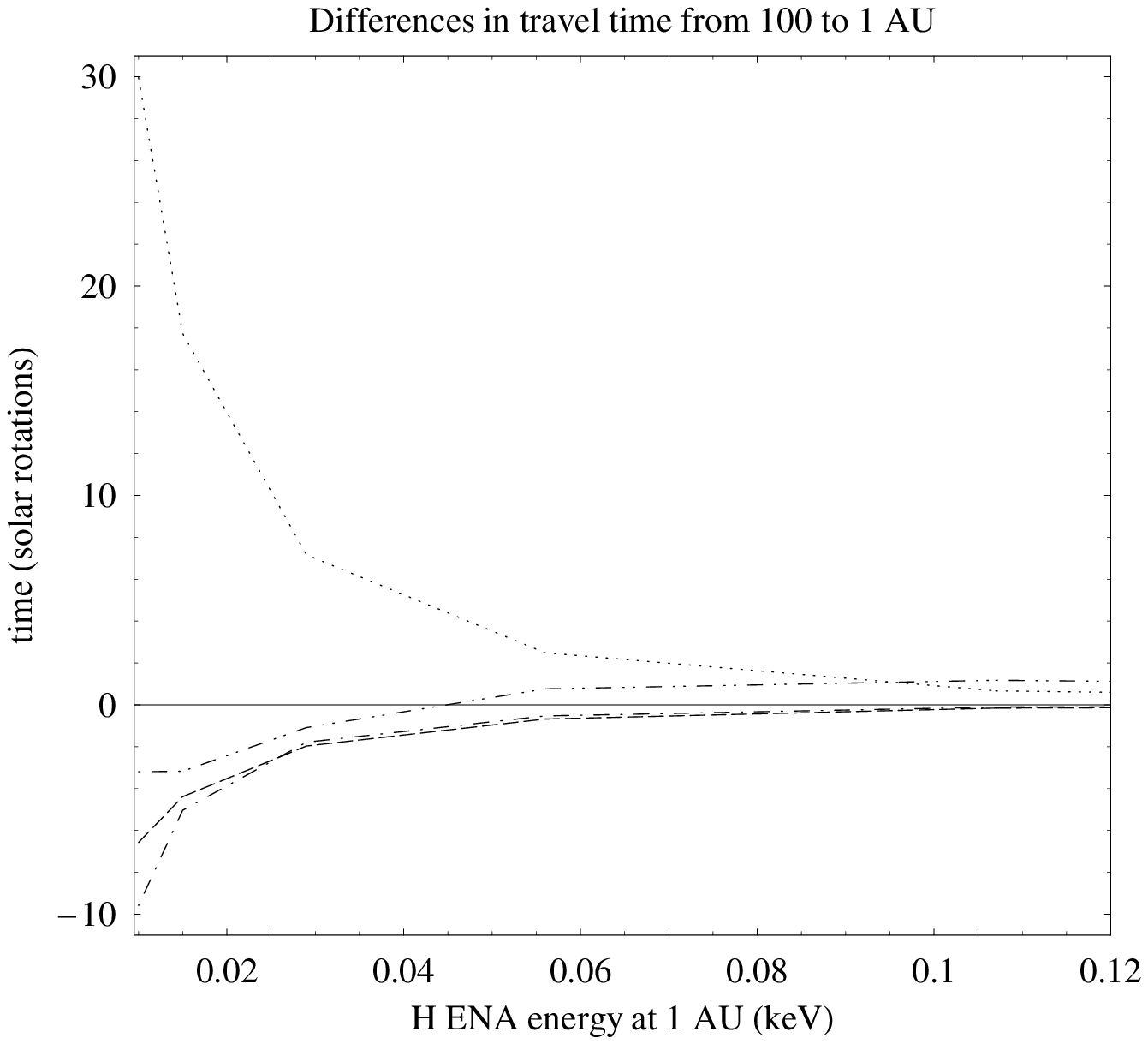}
\caption{{\em Left-hand panel:} Travel times of H ENA from the source region at 100~AU from the Sun to 1~AU as a function of energy at 1~AU. The uppermost solid line corresponds to the travel time with no radiation pressure at all and is shown for reference. The remaining lines correspond to the following cases: (a) solid -- solar minimum, source region in the solar equator plane; (b) dotted -- solar maximum, source region in the solar equatorial plane; (c) dash-dot -- solar minimum, source region at ecliptic poles; (d) dashed -- solar maximum, source region at ecliptic poles; (e) dash-dot-dot (practically overlapping with dash-dot) -- force-free case, when straight-line, constant-speed motion is assumed. {\em Right-hand panel}: Differences in travel time between case (a) -- (b) (dotted line); (a) -- (c) (dash-dot); (b) -- (d) (dashed); and (a) -- (e)  (dash-dot-dot).  }
\label{travelTimePl}
\end{figure*}

\subsection{Travel times of H ENA from source region to 1 AU}

For the high end of the IBEX range of H ENA energies at 1~AU, the travel times from the source region are practically equal to the travel time in the force-free case (straight-line, constant-speed motion). The differences are approximately 1 solar rotation period in duration for energies between 0.1 and 0.2~keV, and for more energetic atoms are even smaller. For $\sim 5$~keV atoms, the travel time is equal to 6 Carrington rotations and for 0.2~keV atoms, it increases to 32.5 rotations. Deviations from the force-free case begin for the 1~AU energies about 0.05~keV and become dramatic towards lower energies, as shown in Fig.~\ref{travelTimePl}. An exception is the purely-ballistic case (which is defined here with no radiation pressure); noticeable differences can be seen already at 0.1~keV of $E_{\mathrm{1~AU}}$ energy and become enormous towards lower energies in the IBEX sensitivity band, for which the travel times of such atoms are twice as long as those of H ENA. The differences in travel time of H ENA due to different values of net solar Lyman-$\alpha$ flux during the solar cycle, range from approximately 10\% at 0.06~keV to 20\% at 0.01~keV. At solar maximum, when the net flux is higher, the atoms detected at a given energy require less time to arrive at 1~AU than during solar minimum, although the radiation pressure is higher. This paradox is explained by the fact that both radiation pressure and solar gravity are effective only relatively close to the Sun. During solar maximum conditions, the atoms become decelerated just prior to their arrival at 1~AU, and consequently spend most of their travel moving much faster than the atoms that arrive during solar minimum, when the atoms follow almost force-free trajectories. This effect, however, exists only for the lowest-energy atoms below $\sim 0.06$~keV at 1~AU.

If the solar Lyman-$\alpha$ flux is indeed lower at the solar poles than at the equator, as discussed in the preceding section, we would expect differences in the travel time between the equatorial and polar regions, which for the lower end  of the IBEX sensitivity band are up to 6 to 8 months, with equatorial atoms arriving more rapidly than the polar ones. 

The atoms that arrive simultaneously at 1~AU with energies within the IBEX sensitivity limit carry information from the source region that pertains to distinctly different epochs, spanning a time interval larger than the solar-cycle period.

\section{Survival probabilities of H ENA}
\begin{figure}
 \centering
\includegraphics[width=8cm]{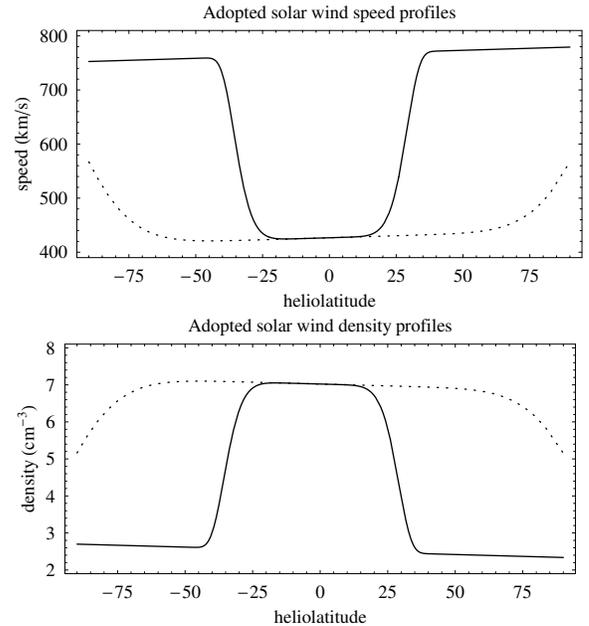}
\caption{Adopted profiles of solar-wind speed (upper panel) and density at 1~AU (lower panel) for solar minimum and maximum conditions (respectively, solid and dotted lines). Corresponding profiles of the flux are shown in the lower panel of Fig.~\ref{absWeightsPl}. }
\label{swprofPl}
\end{figure}
\begin{figure}
\centering
\includegraphics[width=8cm]{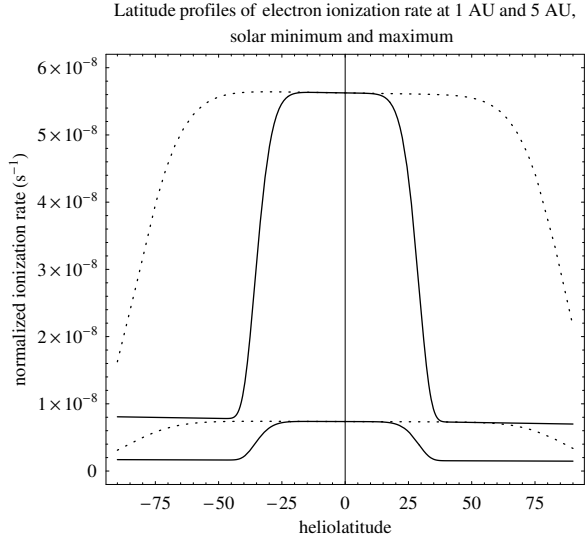}
\caption{Latitudinal profiles of the normalized electron ionization rate $R^2\, \beta_{el}(R)$ at solar minimum (solid lines) and maximum (dotted lines), calculated at 1~AU and at 5~AU. The 5~AU lines are the lower ones.}
\label{betaElLati}
\end{figure}
\begin{figure}
\centering
\includegraphics[width=8cm]{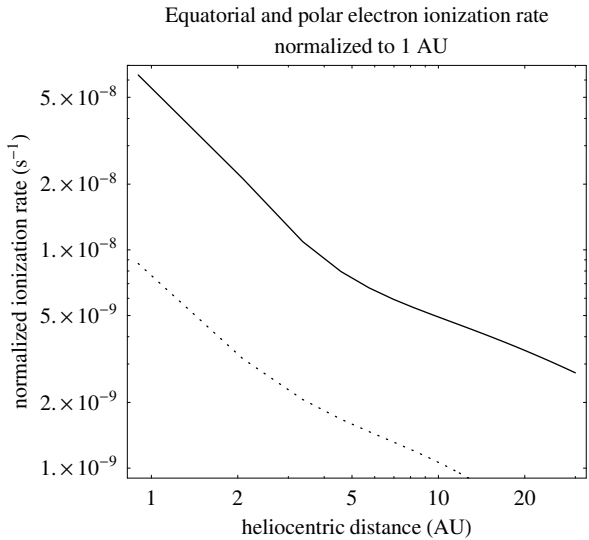}
\caption{Normalized radial profiles $R^2\, \beta_{el}(R)$ of equatorial (solid) and polar (dotted) rates of electron-impact ionization of the incoming H ENA, defined, respectively, by Eq.(\ref{belSlowRate}) and (\ref{belFastRate}).}
\label{belPolarEqtr}
\end{figure}
\begin{figure}
\centering
\includegraphics[width=8cm]{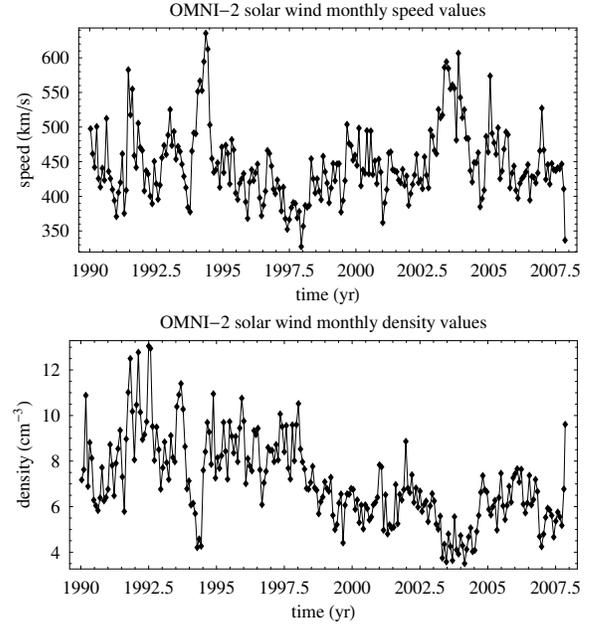}
\caption{Solar wind velocity (upper panel) and density values (lower panel) from the OMNI-2 series, averaged by Carrington rotations.}
\label{omniPl}
\end{figure}
\begin{figure}
 \centering
\includegraphics[width=8cm]{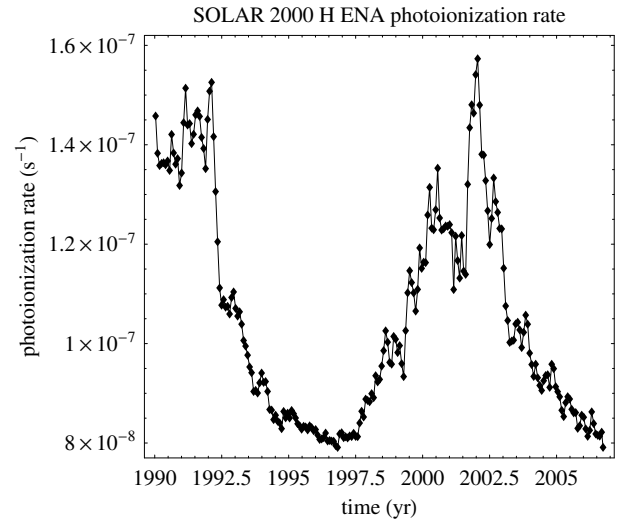}
\caption{Carrington-averaged photoionization rate of H ENA during solar cycle based on SOLAR 2000.}
\label{photoIonPl}
\end{figure}

\subsection{Ionization model used}
The survival probability of H ENA against ionization $w_{\mathrm{ion}}$, between the source region and the Earth orbit can be calculated from the formula:
\begin{equation}
 \label{wionDef}
w_{\mathrm{ion}} = \exp\left[ \int^{t_{\mathrm{source}}}_{t_{1\,\mathrm{AU}}} \beta\left(t, \vec{r}\left(t\right)\right)\, \mathrm{d}t \right],
\end{equation}
where $\beta\left(t, \vec{r}\left(t\right)\right)$ is the ionization rate at time $t$ along the trajectory $\vec{r}\left(t\right)$, $t_{\mathrm{source}}$ is the time of ``launch'' of the atom in the source region, and $t_{1\,\mathrm{AU}}$ is the time of arrival of the atom at 1~AU; the integration goes backwards in time, which results in the absence of the minus sign in the exponent.

The ionization rate $\beta$ is determined by all ionization processes affecting H ENA in the inner heliosphere, i.e. charge exchange with solar wind particles (protons and alphas), collisional ionization by solar-wind electrons, and ionization by solar photons, and was discussed extensively by \citet{bzowski_etal:08a} for neutral interstellar gas in the inner heliosphere. The model of the ionization field relevant for heliospheric H ENA used in the present paper is discussed in greater detail in Appendix A; here, only the most relevant features are pointed out.

To calculate the rate of charge exchange (Eq.~\ref{betaCXRelSpeed}), we need the relative velocities between H ENA and solar wind protons. For trajectories considered in this paper, these velocities vary from an algebraic sum of a H ENA speed $v_{ENA}$ and the solar-wind speed $v_{SW}$ at large distances from the Sun, $v_{\mathrm{rel}} = v_{SW} + v_{ENA}$, to a value appropriate for perihelion, $v_{\mathrm{rel}} = \sqrt{v_{SW}^2 + v_{ENA}^2}$. The change in relative speed occurs over a period of a few months, just before reaching Earth orbit, when changes in the H ENA velocity vector also occur, as discussed in the preceding section. Hence, the rate of charge exchange during the most important time interval of perihelion approach, differs from the usually-adopted $1/r^2$. 

Since the solar wind is bi-modal, that is slow and dense in an equatorial band and fast and rarefied in the polar regions, the field of ionization of the incoming H ENA has latitudinal anisotropy, especially during solar minimum. The range of the slow-wind band expands from solar minimum to maximum with some phase shift between the north and south hemispheres, so that during solar maximum the slow wind persists at practically all latitudes. This fact is taken into account in the model used.

Adopted profiles of solar-wind speed and density are defined by Eqs~(\ref{ec}), (\ref{ed}) and  (\ref{eqphi}) and shown in Fig.~\ref{swprofPl}. The profile widths change in time according to Eq.~(\ref{eqphi}). For solar minimum, the epoch 1995.0 was adopted, and for solar maximum,  epoch 2001.0. The model features some north-south asymmetry, which agrees with the results of observations discussed in Appendix~A.1. It was assumed that the solar-wind speed does not depend on distance from the Sun and that proton density declines as $1/r^2$. 

Since the physical state of electrons in the solar wind depends on the slow/fast regime, the electron-impact ionization rate features a characteristic latitudinal anisotropy, as shown in  Fig.~\ref{betaElLati}. The figure also illustrates the rapid decrease in the electron ionization rate with heliocentric distance, due to cooling of solar-wind electrons. The reduction in the electron ionization rate with solar distance is more rapid than in the case of charge exchange (Fig.~\ref{belPolarEqtr}). The model of electron ionization rate, as presented in greater detail in the Appendix, depends on the density of solar-wind protons at 1~AU $n_{SW}\left(\phi, t\right)$ (Eq.~(\ref{ed})). 

Both the density and speed of the solar wind show strong fluctuations (Fig.~\ref{omniPl}). In the calculations, it was, however, assumed that the equatorial densities and velocities are constant (and their values can be found in Fig.~\ref{swprofPl}); the effect of solar-wind fluctuations on survival probabilities is discussed separately. 

The photoionization rate also shows short-time fluctuations and distinct secular variations that are  correlated with the solar-cycle phase (Fig.~\ref{photoIonPl}). In the calculations, it was assumed that the photoionization rate is invariable and spherically symmetric; for solar minimum at 1~AU, $\beta_{ph,min} = 0.8\times 10^{-7}$~s$^{-1}$, and for solar maximum, $\beta_{ph,max} = 1.4\times 10^{-7}$~s$^{-1}$. It also changes with solar distance, being proportional to $1/r^2$.

\citet{gruntman:90a} pointed out an additional, indirect channel of photoionization losses: a H atom is first excited from ground state to a higher state and then photoionized by solar photons that have a lower energy than needed to perform the ionization directly. This effect increases the photoionization rate by 10\% at 1~AU, but since its efficiency decreases with solar distance far more rapidly than $1/r^2$, it is reduced to just $\sim 1$\% of the direct photoionization rate at 3~AU. Given the uncertainty in the photoionization rate, it can be neglected as an additional mechanism of photoionization losses in the incoming H ENA.

\subsection{Survival probability as function of radiation pressure}

The probability of survival of H ENA traveling to Earth orbit depends on a convolution of the local instantaneous ionization rate along their trajectories with the effects related to their kinematics. Given a fixed ionization field, the probability of survival depends on the effective exposure time to the ionizing effects, which depends on the radiation pressure. This phenomenon is illustrated by the following example. 

\begin{figure}
\centering
\includegraphics[width=8cm]{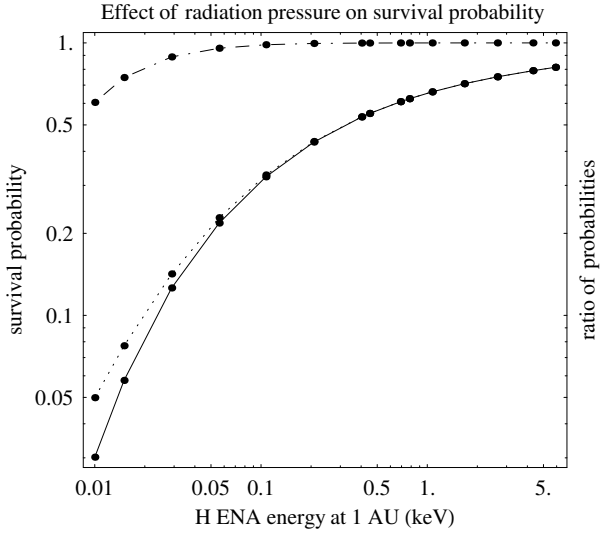}
\caption{Survival probabilities of H ENA for a spherically-symmetric radiation pressure and ionization field. Solid line, left-hand vertical scale: $w_{min}$ for the solar Lyman-$\alpha$ flux equal to $I_{min}$, dotted line: $w_{max}$ for the solar flux equal to $I_{max}$; dash-dot, right-hand vertical scale: ratio of survival probabilities $w_{min}/w_{max}$.}
\label{radPressRole}
\end{figure}

When we assume that both the solar Lyman-$\alpha$ flux and the ionization field are spherically symmetric and that the solar-wind parameters at all latitudes are similar to those at the solar equator (see Figs.~\ref{swprofPl} and  \ref{betaElLati}), then the probability of survival of an H ENA, reaching perihelion 1~AU, does not depend on the arrival direction of the atom. It depends solely on its energy in the source region and on the radiation pressure. This is illustrated in Fig.~\ref{radPressRole}. We show the survival probabilities of an H ENA which reaches perihelion at 1~AU, as a function of energy for solar Lyman-$\alpha$ fluxes $I_{\mathrm{min}}$ and $I_{\mathrm{max}}$. They differ for energies $E_{1\,\mathrm{AU}} < 0.05$~keV, as demonstrated by the dash-dot line in the figure. For atoms of the lowest energies, the differences in survival probability resulting solely from to a change of $I_{\mathrm{tot}}$ from $I_{\mathrm{min}}$ to $I_{\mathrm{max}}$ may be as high as $\sim 40$\%.

In reality, however, the ionization field is not spherically symmetric, and another imprint of the atom dynamics might be expected due to the latitudinal anisotropy of the solar Lyman-$\alpha$ flux. A modulation of the H ENA survival probability as a function of latitude of the approach direction would be expected. Calculations (not shown) suggest, however, that this effect yields a modulation with an amplitude equal to just 5\% of the average value for the lowest energies accessible by IBEX; this amplitude drops down below 1\% for $E_{1\,\mathrm{AU}} = 0.03$~keV, so this effect is negligible. 

\subsection{Contributions of various loss channels}
\begin{figure}
 \centering
\includegraphics[width=8cm]{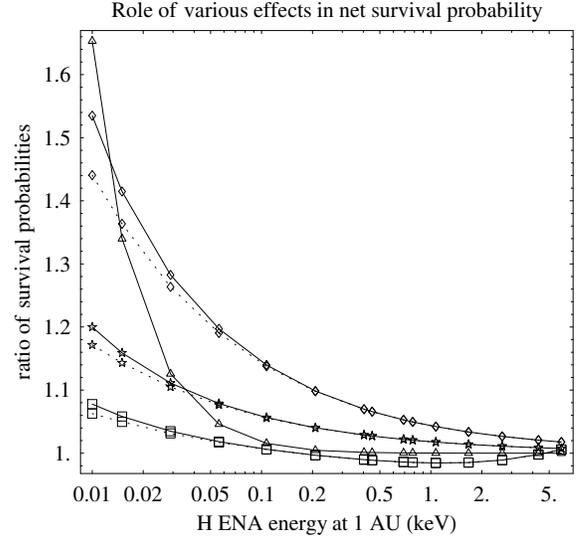}
\caption{Contribution of various parameters to ionization losses of H ENA at Earth orbit. Shown are ratios $w_{ion,x}/w_{ion}$, where $w_{ion}$ is the probability of survival obtained with the use of the full ionization model including charge exchange (with the cross section according to \citet{lindsay_stebbings:05a}), electron ionization, and photoionization, calculated for trajectories in the solar equator plane, and $w_{ion,x}$ is the probability computed along the same trajectories with photoionization switched off (diamonds), with electron ionization switched off (stars), and with a change of the charge exchange cross section from \citet{lindsay_stebbings:05a} to \citet{phaneuf_etal:87} (boxes). Solid lines correspond to the values relevant for solar minimum, broken lines to solar maximum. Additionally shown is the influence of the change of radiation pressure from solar minimum to maximum (triangles) $w_{ion,max}/w_{ion,min}$.}
\label{miscIonRole}
\end{figure}
\begin{figure*}
\centering
\includegraphics[width=16cm]{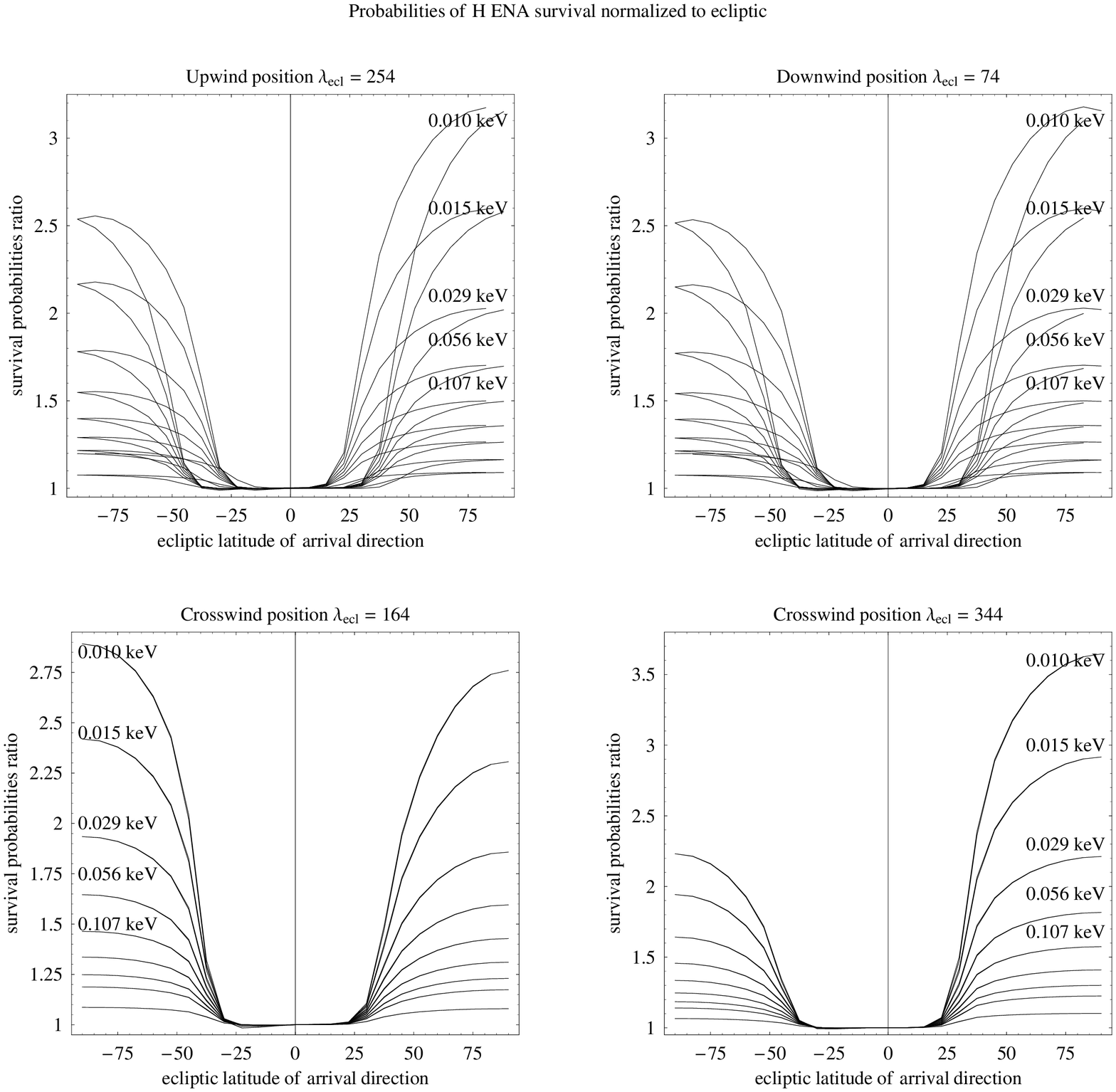}
\caption{Latitudinal profiles of ratios of H ENA survival probabilities $w_{ion}\left(\phi\right)/w_{ion}\left(0\right)$ for various H ENA energies at Earth orbit, shown for solar minimum conditions and for 4 locations at Earth orbit. The energies are indicated in the figure; the profiles without energy labels are for (from top to bottom) 0.208~keV, 0.403~keV, 0.781~keV, and 5.9~keV. The absolute values of probabilities for the reference atoms, arriving from the ecliptic plane, are indicated in Fig.~\ref{radPressRole}. Shown are results of 360\degr scans in the planes perpendicular to the local antisolar vector, parametrized by the ecliptic latitude of the arrival directions of the atoms. }
\label{wionAniso}
\end{figure*}
Contributions from various ionization channels to the net survival probability of H ENA at Earth orbit are presented in Fig.~\ref{miscIonRole}. Shown are ratios $w_{ion,x}/w_{ion}$ calculated for various trajectories, where $w_{ion}$ is the survival probability, obtained from a full model taking into account all three ionization channels, and $w_{ion,x}$ is the probability of survival obtained when a factor ``x'' is switched off. The probabilities were calculated for  trajectories contained in the solar equatorial plane, i.e. fully immersed in the slow solar wind. In such circumstances, the most intensive ionizing factor after charge exchange with solar-wind protons is photoionization, whose share is a strong function of atom energy. In the case of the slowest atoms ($\sim 0.01$~keV), photioinization contributes as much as $\sim 50$\% of the losses, but only $\sim 5$\% for atoms with $E_{1\,\mathrm{AU}} = 6$~keV. During solar minimum, when polar regions are engulfed by the fast wind and the charge exchange rate is much lower than in the slow wind, photoionization however, can be a dominant source of losses of H ENA originating in polar latitudes.

Electron-impact ionization produces up to $\sim 20$\% of the ionization losses of slowest H ENA in the slow solar wind and its contribution decreases to below 10\% at $E_{1\,\mathrm{AU}} \simeq 0.05$~keV. During solar minimum the electron ionization rate is far more anisotropic than the charge exchange rate (cf Fig.~\ref{betaElLati} and the lower panel of Fig.~\ref{absWeightsPl}); we therefore expect that the  contribution of electron ionization to the losses of H ENA approaching Earth orbit from polar latitudes during solar minimum will be lower than in the case of atoms originating in the equatorial latitudes.

The contributions of photoionization and electron ionization to the net losses of H ENA depend weakly on radiation pressure (the change from solar minimum to solar maximum is just a few percent), as illustrated by the dotted lines in Fig.~\ref{miscIonRole}. 

In the case of the slowest H ENA at Earth orbit, the most significant influence on the change in survival probability between solar minimum and maximum is the change in radiation pressure. This is illustrated by the line of triangles in Fig.~\ref{miscIonRole}. The amplitude of change in survival probability of H ENA at transition from solar maximum to solar minimum decreases to below the level of the contribution of photoionization for $E_{1\,\mathrm{AU}} \simeq 0.015$~keV, and below the contribution of electron ionization for $E_{1\,\mathrm{AU}} \simeq 0.03$~keV.

The uncertainty in the cross section for charge exchange (boxes in Fig.~\ref{miscIonRole}) produces an uncertainty of $\sim 10$\% in the estimate of the survival probability of H ENA at the lower end of energies at Earth orbit, and  decreases, with an increase of energy, to the level of a few percent at $\sim 6$~keV.

\subsection{Anisotropy of survival probability }
\begin{figure}
\centering
\includegraphics[width=8.5cm]{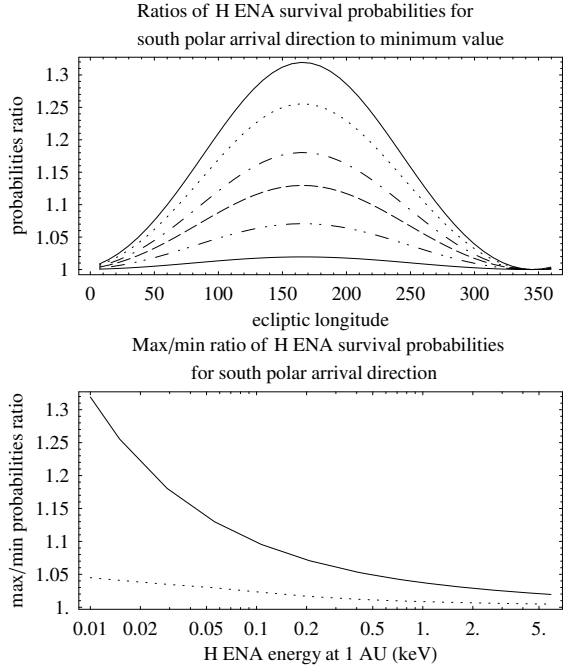}
\caption{{\em Upper panel}: Ratios of survival probabilities of H ENA approaching  Earth orbit from the south ecliptic pole to the lowest probability along the Earth orbit for various H ENA energies at 1~AU: from top to bottom 0.010~keV (solid line), 0.015~keV (dots), 0.029~keV (dash-dot), 0.107~keV (dashed), 0.209~keV (dash-dot-dot), and 5.9~keV (solid). {\em Lower panel}: Ratios of the highest to lowest probability of survival of H ENA originating in the south ecliptic pole region as a function of energy at Earth orbit for solar minimum (solid) and maximum conditions (dotted line).}
\label{yearlyModulation}
\end{figure}
\begin{figure}
 \centering
\includegraphics[width=8cm]{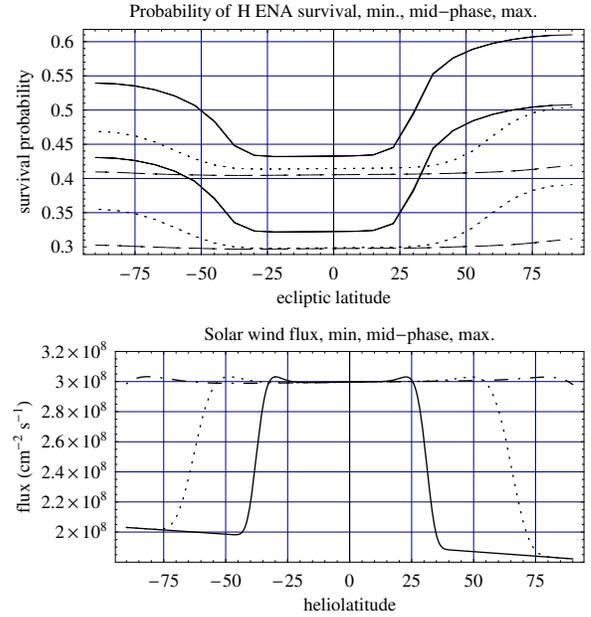}
\caption{{\em Upper panel:} Probabilities of survival of H ENA approaching Earth orbit at ecliptic longitude $344\degr$ as a function of ecliptic latitude of the arrival direction for 1~AU energies equal to 0.107~keV (the lower group of lines) and to 0.209~keV (the upper group of lines) for three phases of solar cycle (from top to bottom in both groups of lines): minimum, intermediate and maximum (respectively: solid, dots, dash). {\em Lower panel:} Adopted solar-wind fluxes as a function of heliolatitude for the three phases of solar cycle (minimum: solid, intermediate: dots, maximum: dash), used to calculate the survival probabilities of the H ENA at Earth orbit shown in the upper panel. The geometry of the upper panel corresponds to the geometry of the lower-right panel of Fig.~\ref{wionAniso}. }
\label{absWeightsPl}
\end{figure}

The probability of survival of H ENA reaching Earth orbit at perihelia of their trajectories exhibits an anisotropy as a function of the latitude of arrival direction, which is a result of combination of the anisotropy of the ionization rate and of geometry effects due to the non-vanishing inclination of Earth orbit to solar equator. In the adopted model, the ionization field has an approximate planar symmetry with respect to the solar equator and IBEX will carry out observations remaining bound in the ecliptic plane, which is inclined to the equator at an angle $\sim 7.25\degr$. The anisotropy is illustrated in Fig.~\ref{wionAniso}. 

Assuming the anisotropic ionization rate as shown in Figs.~\ref{swprofPl} and \ref{betaElLati} and an invariable, anisotropic solar Lyman-$\alpha$ flux (Eq.~(\ref{e4})), the distributions of the survival probability of H ENA of a given energy, reaching Earth orbit at ecliptic longitude $164\degr$ (which is the crosswind point and simultaneously the point in the ecliptic with lowest heliographic latitude $-7.25\degr$; lower-left panel) depend strongly on the latitude of arrival direction. The pole-to-ecliptic probability ratio for the lowest-energetic atoms exceeds 2, but decreases with energy increase, reaching $\sim 1.1$ for the highest-energy atoms. Latitudinal profiles of the survival probability show some north-south asymmetry related to the north-south asymmetry present in the ionization model, but there is no asymmetry in ecliptic longitude (there exists a left-right symmetry). 

A change in the observation point along Earth orbit produces a distortion of this picture, which is entirely due to the shift in the observation point with respect to the solar equator. This effect is shown in the upper-left panel of Fig.~\ref{wionAniso} (a point in the solar equator plane and simultaneously the upwind point of Earth orbit), where one can see -- for the ionization field unchanged -- an appreciable increase in the north-south asymmetry as compared with the previous case and appearance of an asymmetry in ecliptic longitude (the lines in the figure are hysteresis-like, there is no left-right symmetry). Another change in the observation point to the other crosswind point ($\lambda_{ecl} = 344\degr$, a location with the highest heliolatitude, equal to $+7.25\degr$) produces another change in the relative survival probability profiles (lower-right panel): because of the shift with respect to the solar equator, the north-south asymmetry is strengthened considerably and the left-right asymmetry disappears. In the case of the slowest atoms, the north pole-to-ecliptic survival probability ratio appreciably exceeds 3. A transition to the downwind point of the Earth orbit (the other node of solar equator; upper-right panel) of course restores the picture from the upwind point, but the local left and right directions are swapped.

Changes in the survival probabilities of H ENA originating in polar regions are shown in Fig.~\ref{yearlyModulation}. The upper panel of the figure presents the ratio of the survival probabilities $w_{ion,S}\left(\lambda_{ecl}\right)$ of H ENA originating in the south ecliptic pole and reaching Earth orbit at ecliptic longitude $\lambda_{ecl}$, to the lowest probability along the Earth orbit $w_{ion,S}\left(min\right)$. The ratios are shown for various energies at Earth orbit during solar activity minimum. The changes in survival probability shown are solely due to geometric effects. For atoms of the lowest energies at Earth orbit, the yearly amplitude of the ratio $w_{ion,S}\left(\lambda_{ecl}\right)/w_{ion,S}\left(min\right)$ is approximately 30\% and decreases rapidly enough with increasing energy. For energies that exceed $\sim 0.1$~keV, it decreases below 10\% (Fig.~\ref{yearlyModulation}, lower panel). The maximum of survival probability of the H ENA originating in the south ecliptic pole and observed by IBEX will occur at ecliptic longitude $\sim 164\degr$, i.e. at the transition from February to March, and in the case of atoms originating in the north ecliptic pole, the course of survival probability as a function of ecliptic longitude will have a similar shape, but with a phase shift of $180\degr$ (half a year). 

It appears that anisotropies larger than those presented should not be expected. The evolution in the anisotropy of H ENA survival probability for three phases of solar cycle at $\lambda_{ecl} = 344\degr$ for $E_{1\,\mathrm{AU}} \simeq 0.1$~keV and $\sim 0.2$~keV is shown in Fig.~\ref{absWeightsPl}, along with the corresponding profiles of the solar-wind flux. In addition to the solar minimum and maximum phases, as previously, a phase of intermediate solar activity was added, when the band of slow solar wind had expanded to mid-latitudes. For this phase the solar Lyman-$\alpha$ flux $I_{\mathrm{tot}} = 4.7\times10^{11}$~cm$^{-2}$~s$^{-1}$ and the photoionization rate $\beta_{ph} = 1.1\times10^{-7}$~s$^{-1}$ were adopted. The latitudinal profiles of the solar wind flux for these phases of solar activity are shown in the lower panel of Fig.~\ref{absWeightsPl}. The expansion of the latitudinal range of slow solar wind, related to the increase of solar activity, produces a flattening and widening of the profiles of  $w_{ion}\left(\phi\right)/w_{ion}\left(0\right)$, with an almost complete disappearance of anisotropy at solar maximum.

\subsection{Fluctuations of survival probability}
\begin{figure}
 \centering
\includegraphics[width=8cm]{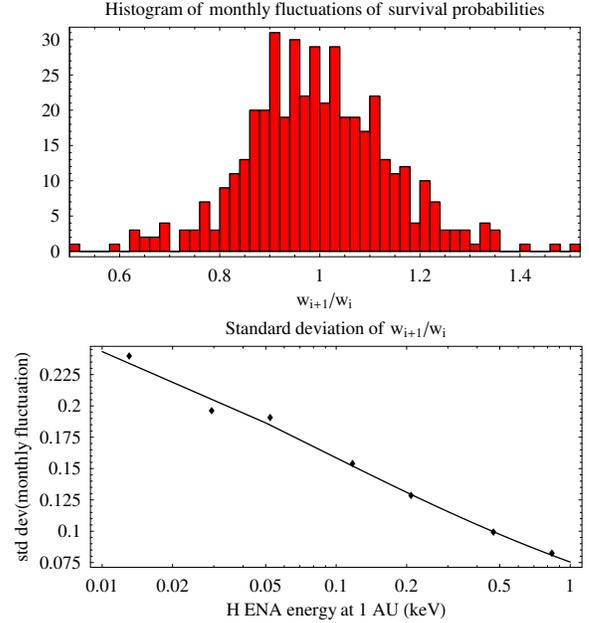}
\caption{{\em Upper panel:} An example histogram of fluctuations of quantity $w_{i+1}/w_i$, where $w_i$ is the probability of survival of H ENA calculated with solar-wind equatorial parameters averaged by Carrington rotation periods for an H ENA energy at Earth orbit equal to 0.1~keV. The solar-wind parameters were taken as the averages of appropriate intervals of the OMNI-2 time series, shown in Fig.~\ref{omniPl}. {\em Lower panel:} Standard deviation of the monthly survival probabilities ratio $w_{i+1}/w_i$ as a function of H ENA energy at Earth orbit (dots) and a plot of the approximation formula specified in Eq.~(\ref{fluctuEq}).}
\label{fluctuPl}
\end{figure}

Because of strong fluctuations in both the solar wind (Fig.~\ref{omniPl}) and the photoionization rate (Fig.~\ref{photoIonPl}), one should expect appreciable fluctuations in the survival probabilities of H ENA at Earth orbit. Available data enable us to estimate these fluctuations for arrival directions contained in the ecliptic plane, but because of insufficient data the estimate for higher latitudes will be less accurate. Since Ulysses observations imply that the fast wind, which is characteristic of high heliolatitudes, is affected by smaller fluctuations than the slow wind, we expect that fluctuations for H ENA originating in the polar regions will be lower than for the ecliptic directions. 

Based on a simplified kinematic model (with a force-free motion of H ENA), the ratios of the H ENA survival probabilities at Earth orbit $w_{i+1}/w_i$ were studied for various energies. For solar-wind conditions that are averaged by $i$-th Carrington rotation, $w_i$ is the probability of survival. Solar-wind parameters from the OMNI-2 time series were used (Fig.~\ref{omniPl}). An example histogram of these ratios for $E_{\mathrm{1\,AU}}=0.1$~keV is shown in the upper panel of Fig.~\ref{fluctuPl}; the distribution is close to normal. The dependence of the standard deviations of these ratios $w_{i+1}/w_i$ on energy, $\sigma_{\mathrm{stdDev}}\left(w_{i+1}/w_i,E_{\mathrm{1\,AU}}\right)$, is illustrated by the dotted symbols in the lower panel of Fig.~\ref{fluctuPl}. This dependence can be approximated by the formula: 
\begin{equation}
 \label{fluctuEq}
\sigma_{\mathrm{stdDev}}\left(w_{i+1}/w_i,E_{\mathrm{1\,AU}}\right)=0.075 \exp\left[-0.03 \ln^2\left(E_{\mathrm{1\,AU}}\right)\right]\,E_{\mathrm{1\,AU}}^{-0.39}, 
\end{equation}
with energy $E_{\mathrm{1\,AU}}$ expressed in keV, and is shown in the lower panel of the figure with the solid line. This suggests that the fluctuations $\delta\,w$ in the survival probability at Earth orbit, due to fluctuations in the ionizing factors, should range $0.2\, w_{\mathrm exp}$, where $w_{\mathrm exp}$ is the expected probability value at the lower end of the IBEX energy band, to negligible values ($\delta\,w \simeq 0$ in the case of H ENA from the high limit of this band. For H ENA of energies expected from the heliosheath 0.1 -- 0.2~keV, the fluctuations are on the order of $\sim 0.2\, w_{\rm exp} - 0.15\, w_{\rm exp}$, which is smaller than the amplitude of the latutidinal modulation related to the solar-wind anisotropy (Figs.~\ref{wionAniso} and \ref{absWeightsPl}). 

\section{Summary and conclusions}

We investigate modifications of H ENA energy at Earth orbit with respect to their energy at 100~AU from the Sun and their probabilities of survival against losses induced by local heliospheric ionization processes. The range of energy at Earth orbit that we have studied corresponds to the sensitivity range of the planned NASA SMEX mission IBEX, with emphasis on the H ENA energies expected from the inner heliosheath (mostly 0.1 -- 0.2~keV, up to $\sim 1$~keV). Realistic, observation-based models of both ionization processes and radiation pressure have been used. The ionization model reproduces the time-dependent spatial anisotropy of the solar wind, and the radiation-pressure model takes into account the dependence of $\mu$ on both the radial velocities of the atoms and on the heliolatitude (Figs.~\ref{lyaPlot}, \ref{swprofPl}, \ref{betaElLati}, \ref{omniPl}, and \ref{photoIonPl} and Eqs~(\ref{equ2}), (\ref{e3}), and (\ref{e4})). 

Based on the results of our numerical simulations,  we propose that, due to radiation pressure, H ENA reach Earth orbit without a significant changes in energy and direction apart from the atoms with energies below 0.1~keV during high solar activity (Figs.~\ref{sourceEnClipped}, \ref{deflectionPlot}). The probability of H ENA survival is a function of the solar-cycle phase, the ecliptic longitude of the location at Earth orbit, and the ecliptic latitude of the approach direction. Especially important in this respect is the latitudinal range of the slow solar-wind region (Figs.~\ref{wionAniso} and \ref{absWeightsPl}). For all cases considered, the time of flight from the source region at 100~AU to Earth orbit is equal to the force-free travel time, to an accuracy of 2 solar rotation periods, down to 1~AU energies $\sim 0.05$~keV.

The survival probability of H ENA declines rapidly with decreasing energy at Earth orbit from $\sim 0.8$ for $E_{\mathrm{1\,AU}} = 6$~ keV to $\sim 0.4$ for $E_{\mathrm{1\,AU}} = 0.2$~keV and $\sim 0.3$ for $E_{\mathrm{1\,AU}} = 0.1$~keV for H ENA approach directions within the ecliptic plane (Fig.~\ref{radPressRole}). A yearly modulation of the survival probability of H ENA originating in the polar regions, is expected due to the yearly movement of the  Earth-bound detector in heliolatitude. The yearly amplitude of the maximum-to-minimum probability ratio decreases with increasing  energy at Earth orbit; for $E_{\mathrm{1\,AU}} \simeq 0.1$~keV the ratio is equal to $\sim 1.15$, and for 0.2 keV to $\sim 1.1$ (Fig.~\ref{yearlyModulation}). The courses of the modulation for the north and south poles are in antiphase. 

Apart from a short time interval at solar maximum, one also expects an appreciable modulation of the survival probability as a function of heliolatitude (Figs.~\ref{wionAniso} and \ref{absWeightsPl}). During solar minimum in the case of H ENA with $E_{\mathrm{1\,AU}} = 0.1$~keV the amplitude of this modulation should not exceed 1.6, 1.3 for $E_{\mathrm{1\,AU}} = 0.2$~keV and 1.1 for $E_{\mathrm{1\,AU}} = 6$~keV. Appreciable left-right and north-south asymmetries are expected too, changing with position at Earth orbit (Figs.~\ref{wionAniso} and \ref{absWeightsPl}), as well as a distinct correlation of the survival probability profiles with the latitude structure of solar wind. The amplitude of the latter modulation decreases with widening of the slow solar wind region (Fig.~\ref{absWeightsPl}). 

Because of fluctuations of the solar wind and of solar EUV radiation at time scales shorter than solar rotation period fluctuations of the H ENA survival probability at Earth orbit are expected. Based on the solar wind observations at 1~AU and on the simulations discussed in the paper a phenomenology function expressing the amplitude of these fluctuations with respect to the survival probability as a function of the H ENA energy at Earth has been found (Eq.(\ref{fluctuEq})): it decreases with an increase of energy and becomes lower than 15\% for $E_{\mathrm{1\,AU}}\simeq 0.1$~keV (Fig.~\ref{fluctuPl}). 

We have also studied the contributions of various factors affecting the losses of H ENA traveling from the source region to Earth orbit. The uncertainty in the charge-exchange cross-section yields uncertainty in the survival probability approximately of a few percent, much lower than the amplitude of the probability fluctuations due to fluctuations in the solar wind. The strength of radiation pressure affects the survival probability on the level of $\sim 20$\% for $E_{\mathrm{1\,AU}}\simeq0.1$~keV and $\sim 15$\% for $E_{\mathrm{1\,AU}}\simeq0.2$~keV. Photoionization contributes 20\% and 15\% of the losses at the energies $E_{\rm 1\,AU} \simeq 0.1$~keV and $E_{\rm 1\,AU} \simeq  0.2$~keV, respectively, and far more for approach directions from high latitudes. Electron ionization for energies above 0.1~keV contributes less than 10\% of the losses (Fig.~\ref{miscIonRole}).

All modulation factors discussed increase appreciably -- although to varying degrees -- with a decrease in H ENA energy at Earth orbit. They should, however, disappear during solar maximum, apart from those related to solar-wind fluctuations, when the large-scale structure of the solar wind is spherically symmetric. 

The analysis presented in this paper has neglected the observational effect of a proper motion of the detector, which will be the subject of future studies. Based on results obtained thus far it is concluded that the most important elements needed for a successful interpretation of H ENA measurements, such as those planned for the forthcoming IBEX mission, are monitoring of the local solar wind, the latutudinal structure of the solar wind, and the solar Lyman-$\alpha$ flux. The solar EUV flux in the spectral range responsible for photoionization of hydrogen is also important, and especially its dependence on heliolatitude. Taking into account the local effects of H ENA transport from source to Earth orbit, we expect that IBEX should be able to discover departures from symmetry in the heliospheric interface of approximately a few percent.

\begin{appendix}
\section{Ionization processes destroying heliospheric H ENA}
\subsection{Charge exchange with solar-wind protons and alpha particles}

\begin{figure}
 \centering
\includegraphics[width=8cm]{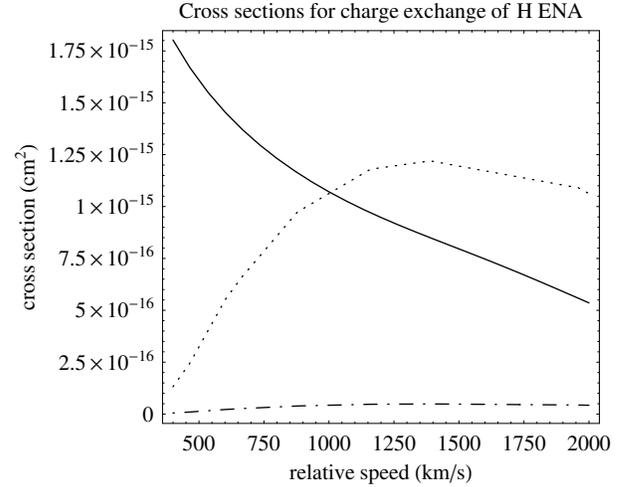}
\caption{Cross sections for charge exchange between H ENA and protons (solid line) and alpha particles (dotted line) for collision speeds typical of the inner heliosphere. Broken line is the alpha charge exchange rate multiplied by a typical abundance of alphas with respect to protons, which is assumed to be 0.04. }
\label{sigcxPl}
\end{figure}
The local rate of charge exchange between a H ENA traveling with velocity $\vec{v}_{ENA}$ and a proton or alpha particle traveling with velocity $\vec{v}_{p}$ and a member of solar-wind population described by a local distribution function $f_{p}\left(\vec{v}_{p}\right)$, is defined by the formula:
\begin{equation}
\label{betaCXgeneral}
\beta _{\mathrm{cx}}=\int \sigma _{\mathrm{cx}}\left(\left|\vec{v}_{ENA}-\vec{v}_{p}\right|\right) \left|\vec{v}_{ENA}-\vec{v}_{p}\right| f_{p}\left(\vec{v}_{p}\right)\,\mathrm{d}\vec{v}_{p}, 
\end{equation}
where $\sigma _{\mathrm{cx}}$ is the charge-exchange cross-section, which is a function of the relative velocity between the colliding particles $\vec{v}_{\mathrm{rel}} \equiv \vec{v}_{ENA}-\vec{v}_{p}$. Since the solar wind is hypersonic, the thermal spread of its positively-charged particles can be neglected. The distribution function $f_p$ is then replaced with a delta function and with the substitution 
\begin{equation}
\label{vrelDef}
v_{\mathrm{rel}} \equiv  \left|\vec{v}_{\mathrm{rel}}\right| = \left|\vec{v}_{H} - \vec{v}_{\mathrm{SW}} \right|,
\end{equation}
 where $\vec{v}_{\mathrm{SW}}$ is the local solar-wind velocity vector, Eq.~(\ref{betaCXgeneral}) simplifies to the form:
\begin{equation}
 \label{betaCXRelSpeed}
\beta _{\mathrm{cx}} = \sigma _{\mathrm{cx}}\left(v_{\mathrm{rel}}\right)\, v_{\mathrm{rel}} \,n_{\mathrm{SW}}
\end{equation}
where $n_{\mathrm{SW}}$ is the local density of solar wind~protons.

Since the solar-wind velocity in the inner heliosphere can be treated as purely radial and the incoming H ENA travel almost radially towards the Sun, we can determine approximate limits of the relative velocities between the colliding solar-wind ions and the H ENA that at 1~AU have energies within the IBEX sensitivity limits: from $\sim 400$~km/s for the slowest ENA and slow solar wind to $\sim 2000$~km/s for the fastest ENA and fast solar wind. This allows us to assess the cross sections for the charge exchange of H ENA with solar-wind protons and alpha particles: for a H + He$^{++}$ collision the cross section varies from $1.3\times10^{-16}$~cm$^2$ for the low-energy collision to $10^{-15}$~cm$^2$ for the high-energy collision \citep{phaneuf_etal:87} and for the H + H$^+$ collision respectively from $1.8\times 10^{-15}$~cm$^2$ to $5\times 10^{-16}$~cm$^2$ \citep{lindsay_stebbings:05a}. The cross sections for charge exchange with protons are larger than the cross sections for charge exchange with alphas only for relative velocities below $\sim 1000$~km/s (Fig.~\ref{sigcxPl}): since the abundance of solar-wind alpha particles with respect to protons is, however, typically just a few percent \citep{vonsteiger_etal:00a}, the net contribution of the charge exchange with alphas is negligible for slow H ENA and up to 10\% for the fast wind and fastest ENA. 

The precise value of the charge exchange cross-section has been a subject of debate. Fortunately for the topic of this paper, estimates by various authors \citep{barnett_etal:90, lindsay_stebbings:05a} appear to have converged within $\sim 10$\% for the collision speeds relevant to H ENA charge exchange with solar wind protons (apart from the highest energy wing, for which they increase to $\sim 20$\%). More details about this issue can be found in \citet{fahr_etal:07a}. In the simulations discussed in this paper, the cross section recommended by \citet{lindsay_stebbings:05a} was used.

The solar wind is known to be anisotropic, consisting of an equatorial band of slower expansion of a denser plasma (interleaved with high-speed streams) and polar caps of a steady, rarefied, but much faster wind. The latitudinal range of the jagged boundaries of the slow wind band vary during the solar cycle, being closest to the equator during solar minimum and expanding gradually towards the poles with an increasing level of solar activity, before engulfing the poles during the solar-maximum phase. The north and south hemispheres appear to be shifted in phase by a year or two. 

Details of this evolution are unknown. Latitudinal structure of the solar-wind was measured in situ by Ulysses, mostly during fast latitude scans \citep{phillips_etal:95b, mccomas_etal:99, mccomas_etal:00b, mccomas_etal:03a}. These measurements, although highly accurate, provide only point-like data on density and velocity without a global snapshot picture. In particular, they provide little information about the longitudinal pattern of the fast/slow wind boundary. 

Another technique to study the solar wind structure is mapping the solar wind indirectly, by analysis of the solar Lyman-$\alpha$ radiation backscattered off the heliospheric neutral hydrogen \citep{bertaux_etal:95}. The dense and slow equatorial solar wind is able to ionize the neutral interstellar hydrogen inside the heliosphere more efficiently than the fast but rarefied wind in the polar regions, and thus carves a trough in the gas distribution, seen as a dimmer band in the global maps of heliospheric glow, referred to as the heliospheric groove \citep[e.g.][]{kyrola_etal:98, bertaux_etal:99}. \citet{bzowski:03} proposed a quantitative model to infer the latitudinal span of the equatorial band of the slow wind from the observations of the latitudinal range of the groove, employed by \citet{bzowski_etal:03} to determine the evolution of the slow-wind region from solar minimum in 1996 to solar maximum in 2001. \citet{bzowski_etal:08a} pointed out that there is a correlation between the latitudinal span of coronal holes and the boundaries of the slow wind region. Similarly, \citet{quemerais_etal:07a} demonstrated a correlation between coronal features observed in white light and the structure of the solar wind. Based on such correlations and on remote-sensing information, \citet{bzowski_etal:08a} suggested a phenomenology, proxy-based model for the evolution of structure in the solar wind, which forms the basis of construct a of solar-wind model constructed to calculate the charge-exchange losses discussed in this paper.

The Lyman-$\alpha$ remote-sensing technique is only able to recover the structure in the ionization rate of the low-energy heliospheric hydrogen, which is correlated with the solar-wind flux. Thus, it is able to yield a global structure of the solar wind over an extended interval of time, but is unable to recover separately the solar-wind speed and velocity, which are essential to calculate the charge exchange rate of the incoming H ENA (see Eq.~(\ref{betaCXRelSpeed}) and (\ref{vrelDef})). 

In addition to Ulysses in situ and SWAN remote-sensing observations we, fortunately, also have continuous monitoring of the solar wind in the ecliptic plane. The data retrieved from various experiments were cross-calibrated and compiled into the OMNI time series \citep{king_papitashvili:05}, which covers a few solar cycles until present (Fig.~\ref{omniPl}). As one can see in the figure, effects of the level of solar activity are not so obvious as they were in the case of the solar EUV output shown in Figs.~\ref{lyaPlot} and \ref{photoIonPl}. We note a downward trend in solar-wind density, which initiated about the previous solar minimum and appears to continue until the present; a clear, periodic signal appears, however, to be missing. Velocity shows some correlation with solar activity, increasing during solar maximum and decreasing during solar minimum. The amplitude of these variations, however, is only about 10\%, so the flux is mostly governed by the density changes, and decreases steadily during the past solar cycle. 

With these data sets and observations in hand, we attempt to construct an evolutionary model of latitudinal density and velocity structure of the solar wind. We develop the following model by assuming that (i) the values of the fast solar-wind density and velocity at the poles are invariable (although the north- and south-pole quantities may differ from each other), that (ii) the fast/slow wind boundaries are parallel to the solar equator (not exactly true, but close to the truth when averaged over all solar longitudes) and equal to the values measured by Ulysses during the fast latitude scan, and that (iii) the equatorial values correspond to the values provided by the OMNI series (which agree with Ulysses values at its passages through ecliptic), and adopting the variations in the north and south boundaries of the slow wind band as inferred from the SWAN Lyman-$\alpha$ mapping and the coronal proxies.

The latitudinal profile of the solar-wind speed is defined by the formula:
\begin{eqnarray}
\label{ec}
v_{\mathrm{SW}} \left(\phi ,t\right)&=&\left(v_{\mathrm{SW,pol}} + \delta_{ v}\, \phi \right) + \left(v _{\mathrm{SW,eqtr}}\left(t\right)-v_{\mathrm{SW,pol}}\right)\\  
& \times &\exp \left[-\ln  2 \left(\frac{2 \phi  - \phi_{N}\left(t\right)- \phi_{S}\left(t\right)}{\phi_{N}\left(t\right)-\phi_{S}\left(t\right)}\right)^N\right],\nonumber
\end{eqnarray}
where $\phi $ is heliographic latitude and $N$ is a shape factor, which in the present study was adopted to be $N=8$; $v_{\mathrm{SW,pol}}$ is the average solar-wind velocity at the poles and the term $\left(v_{\mathrm{SW,pol}}+ \delta_{v}\, \phi \right)$ describes the north-south asymmetry of the polar velocities; the term $\left(v_{\mathrm{SW,pol}} + \delta_{v}\, \phi \right)+\left(v_{\mathrm{SW,eqtr}}\left(t\right)-v_{\mathrm{SW,pol}}\right)$ for $\phi = 0$ corresponds to the solar wind speed at solar equator; and the term $\exp\left[-\ln 2 \left(\frac{2 \phi - \phi_{N} - \phi_{S}}{\phi_{N}-\phi_{S}}\right)^N\right]$ describes the latitudinal dependence of the velocity. 

The density profile $n_{\mathrm{SW}} \left(\phi ,t\right)$ is defined by a similar formula: 
\begin{eqnarray}
\label{ed}
R^2 \,n_{\mathrm{SW}} \left(\phi ,t\right)&=&\left(n_{\mathrm{SW,pol}} + \delta_{ n}\, \phi \right) + \left(n _{\mathrm{SW,eqtr}}\left(t\right)-n_{\mathrm{SW,pol}}\right)\\  
& \times &\exp \left[-\ln  2 \left(\frac{2 \phi  - \phi_{N}\left(t\right)- \phi_{S}\left(t\right)}{\phi_{N}\left(t\right)-\phi_{S}\left(t\right)}\right)^N\right],\nonumber
\end{eqnarray}
where $R = r/r_E$ ($r_E \equiv 1$~AU). The parameters in Eqs (\ref{ec}) and (\ref{ed}) were fitted to the Ulysses data from the first fast latitude scan and the results are the following: $v_{\mathrm{SW,pol}} = 766.0$~km/s, $\delta_v = 0.14822$~km/s/deg, $n_{\mathrm{SW,pol}} = 2.5180$~cm$^{-3}$, $\delta_n = -2.0009\cdot10^{-3}$~cm$^{-3}$/deg. The evolution of the north and south boundaries of the equatorial slow-wind region is adopted from \citet{bzowski_etal:08a}:
\begin{equation}
\label{eqphi}
\phi_{N,S}\left(t\right) = \phi_{0\,N,S} + \phi_{1\,N/S}\, \exp\left[-\cos^3 \left(\omega_{\phi\,N,S}t \right) \right],
\end{equation}
where $\phi_{0,N} = 36.4\degr$, $\phi_{1,N} = -22.2\degr$, $\omega_{\phi,N} = 0.58251$, $\phi_{0,S} = -40.600\degr$, $\phi_{1,S} = 20.0\degr$, $\omega_{\phi_S} = 0.58226$. For realistic modeling, the values of parameters $n_{\mathrm{eqtr}}$, and $v_{\mathrm{eqtr}}$ for a given time $t$ (which must be expressed in decimal years) should be adopted from the OMNI time series (Fig.~\ref{omniPl}); in the fits to the first Fast Latitude Scan by Ulysses used throughout this paper, they were equal to 7.02~cm$^{-3}$ and 426.7~km/s, respectively. The density and velocity profiles returned by the model for solar minimum and maximum are shown in Fig.~\ref{swprofPl} and the flux in the lower panel of Fig.~\ref{absWeightsPl}. 

\subsection{Photoionization}

The rate of photoionization of heliospheric hydrogen is surprisingly poorly understood. To measure the value $\beta_{\mathrm{ph}}$, we must integrate solar spectral flux $F\left(\lambda\right)$ in the energy range above the H photoionization threshold at 13.59~eV ($\lambda_0 = 91.175$~nm), multiplied with the energy-specific photoionization cross-section $\sigma_{\mathrm{ph}}\left(\lambda\right)$:
\begin{equation}
 \label{eqPhotoIon}
\beta_{\mathrm{ph}} = \int_{\lambda_0}^{0} F\left(\lambda\right) \,\sigma_{\mathrm{ph}}\left(\lambda\right)\,\mathrm{d}\lambda
\end{equation}
Details can be found, e.g., in \citet{ogawa_etal:95}. The solar spectrum in the relevant frequency range consists of a continuum, which decreases rapidly with decreasing wavelength, and of a flickering line-component, which is responsible mostly for solar cycle-related variations in the net photoionization rate. Hence, monitoring the neutral H photoionization rate requires a continuous monitoring of the solar spectrum shortwards of the photoionization threshold. While such measurements were  performed by SOHO, there exists a spectral gap in the coverage just short of the ionization threshold, which must be filled with proxies. In the past, a very simple proxy scheme based on the solar 10.7~cm radio flux was in use \citep{rucinski_etal:96a, bzowski:01a}. Recently, a more ingenuous proxy scheme was developed within the framework of the SOLAR 2000 model \citep{tobiska_etal:00c}, which was adopted in the present paper. 

The photoionization rate shows a strong correlation with the solar Lyman-$\alpha$ flux. It varies distinctly during the solar cycle, with a mean value equal to about $1.1\times10^{-7}$~s$^{-1}$ and an amplitude of approximately 30\% (Fig.~\ref{photoIonPl}). Thus, the photoionization rate is equal to about 20\% of the equatorial charge exchange rate. It had been often assumed to be spherically symmetric, but studies by \citet{witte:04}, concerning the photoionization rate of neutral heliospheric helium, supported by \citet{auchere_etal:05c}, imply that the hydrogen photoionization rate should HAVE a heliolatitude anisotropy, which varies during the solar cycle. Due to insufficient data, this effect has been omitted from our present analysis. 

Since the inner heliosphere is almost perfectly transparent to solar EUV radiation of wavelengths shorter than the hydrogen ionization threshold, the H photoionization rate decreases with solar distance as $1/r^2$. 

\subsection{Electron impact ionization}

Construction of a model of electron-impact ionization in the incoming H ENA is hampered by the fact that the behavior of the  electron fluid in the solar wind is not fully understood. It is known that their local distribution function can be decomposed into 3 components: a warm core, a hot halo (both approximated by Maxwellians), and a fluctuating spur, stretched along the local magnetic-field direction \citep{pilipp_etal:87c}. Electron density can be taken from quasi-neutrality and continuity conditions of the solar wind and hence can be adopted as equal to the local proton density + 2 x alpha density. The temperature in the region of interest (i.e. outside 1~AU at all heliolatitudes) is more uncertain. From measurements performed by Ulysses, it appears that it differs from that of protons and that its radial profile can be described by power laws with exponents in the range between isothermal and adiabatic values. The profiles in the fast wind differ from those in the slow wind. The content of the halo population is a function of heliocentric distance and of the solar-wind regime (fast/slow). These findings were recapitulated by \citet{maksimovic_etal:00a}. The measurement of the following quantities are, however, unreliable: the exponents of the temperature power laws, the balance between the core and halo populations, and their evolution during the solar cycle. The construction of an accurate model of electron ionization is therefore presently impossible.
Fortunately, electron ionization is a minor loss process in the incoming H ENA and thus an approximate model is sufficient.

The electron ionization rate of an incoming ENA can be calculated from the formula:
\begin{equation}
\label{beta-el} 
\beta_{\mathrm{el}} = \int_{v_{\mathrm{lim}}}^{\infty} f_e\left(v_{\mathrm{rel}}\right)\,\sigma_e\left(v_{\mathrm{rel}}\right)\, v_{\mathrm{rel}}\, \mathrm{d}^3 v_{\mathrm{rel}}
\end{equation}
where $\sigma_e$ is the cross section for electron-impact ionization \citep{lotz:67}, $f_e$ is the local electron distribution function, $v_{\mathrm{rel}}$ is the relative speed between the incoming ENA and a solar-wind electron, and $v_{\mathrm{lim}}$ is the velocity corresponding to the kinetic energy which is equal to the hydrogen ionization potential 13.59~eV. At electron temperatures of the order of $10^4 - 10^5$~K, the thermal spread in the electron distribution function is large in comparison with the relative velocities between the incoming H ENAs and individual electrons, and $v_{\mathrm rel}$ in Eq.~(\ref{beta-el}) can be superseded by the electron specific velocity $v$. Following \citet{rucinski_fahr:89}, the electron-impact ionization rate can be calculated by assuming that the local distribution function of electrons is composed of two Maxwellian functions (core and halo), parametrized respectively by core and halo densities $n_c$ and $n_h$ and core and halo temperatures $T_c$ and $T_h$. 

These calculations were performed independently for the fast- and slow-wind regimes using parameters for the slow wind adopted from \citet{scime_etal:94}:
\begin{eqnarray}
\label{elSlowWind}
T_c & = & 1.3\cdot 10^5\,r^{-0.85} \nonumber\\
T_h & = & 9.2\cdot 10^5\,r^{-0.38} \\
\xi_{ch} & \equiv & n_h/n_c = 0.06\,r^{-0.25} \nonumber 
\end{eqnarray}
and, for the fast-wind regime, from \citet{issautier_etal:98} and \citet{maksimovic_etal:00a}:
\begin{eqnarray}
 \label{elFastWind}
T_c & = & 7.5\cdot 10^4\,r^{-0.64} \nonumber \\
T_h/T_c & = & 13.57 \\
\xi_{ch}  & = & 0.03 \nonumber
\end{eqnarray}
In both solar wind regimes, the core and halo densities $n_c$ and $n_h$ are calculated from the equations:
\begin{eqnarray}
\label{elCoreDens}
 n_c & = & \frac{1 + 2 \xi_{\alpha}}{1 + \xi_{ch}}\, n_p \nonumber \\
n_h & = & \xi_{ch}\, n_c
\end{eqnarray}
with the alpha abundance $\xi_{\alpha} = 0.04$, which is identical in both fast and slow wind regimes. $\xi_{ch}$ is the halo-to-core density ratio defined for slow and fast wind in Eq.~(\ref{elSlowWind}) and (\ref{elFastWind}), respectively, and $n_p$ is the proton density adopted from the density model discussed earlier in this paper. 

Based on these equations, calculations of the electron ionization rate as a function of heliocentric distance were performed (separately for the fast and slow modes) and the results were approximated by the formula:
\begin{equation}
\label{elradialAppr}
 R^2 \beta_{el}\left(R, n_p\right) = n_p\, \exp\left[\frac{c\,\left(\ln R\right)^2 + b \ln R+a}{g \left( \ln
   R\right)^3+f \left(\ln R\right)^2+e \ln R + d} \right].
\end{equation}
The corresponding radial profiles of the electron-impact rate for the slow and fast wind are presented in a numerically-optimized form in Eq.~(\ref{belSlowRate}) and (\ref{belFastRate}), respectively:
\begin{eqnarray}
\label{belSlowRate}
\beta_{\mathrm{el,s}}\left(R, n_p\right) & = & \frac{n_p}{R^2}\,\mathrm{e}^{\frac{\displaystyle \ln R \left(541.69 \ln R - 1061.32 \right) + 1584.32}{\displaystyle \left(\ln R- 29.17\right)\left(\left(\ln R - 2.02\right) \ln R + 2.91 \right)}} \\
\label{belFastRate}
\beta_{\mathrm{el,f}}\left(R, n_p\right) & = & \frac{n_p}{R^2}\,\mathrm{e}^{\frac{\displaystyle \ln R \left(348.73 \ln R -917.39\right) + 2138.05}{\displaystyle \left(\ln R - 18.97 \right)\left(\left( \ln R - 2.53\right) \ln R + 5.74\right)}}
\end{eqnarray}
As is evident in these formulae, the electron-impact ionization rates are parametrized by local proton densities $n_p$ normalized to 1~AU and are fixed functions of heliocentric distance, which differ appreciably from the $1/R^2$ profiles that are typical of the solar wind flux and photoionization rate, as shown in Fig.~\ref{belPolarEqtr}. 

The global time- and latitude-dependent model of the electron impact ionization rate is constructed by analogy with the proton-density model:
\begin{eqnarray}
\label{betaElGlobal}
\beta_{\mathrm{el}}\left(R, \phi, t\right) &=& \left(\beta_{\mathrm{el,pol}} + \phi\, \delta_e\right)  + \left(\beta_{\mathrm{el, eqtr}} - \beta_{\mathrm{el,pol}}\right)\\
 & \times &\exp \left[-\ln  2 \left(\frac{2 \phi  - \phi_{N}\left(t\right)- \phi_{S}\left(t\right)}{\phi_{N}\left(t\right)-\phi_{S}\left(t\right)}\right)^N\right]\nonumber
\end{eqnarray}
where $\phi_N\left(t\right), \phi_S\left(t\right)$ are defined in Eq.(\ref{eqphi}), and 
\begin{eqnarray}
\label{betaEqrtpol}
\beta_{\mathrm{el, eqtr}} & = & \beta_{\mathrm{el,s}}\left(R, n_{\mathrm{eqtr}}\left(t\right)\right) \nonumber \\
\beta_{\mathrm{el,pol}} & = & \beta_{\mathrm{el,f}}\left(R, n_{\mathrm{pol}}\left(t\right)\right) \\
\delta_e & = & \beta_{\mathrm{el,f}}\left(R, \delta_n\right)  \nonumber
\end{eqnarray}
with solar-wind related densities $n_{\mathrm{eqtr}}, n_{\mathrm{pol}}$ taken from the model of solar-wind density described by Eq.~(\ref{ed}), and the models of the slow and fast wind, indicated by Eq.~(\ref{belSlowRate}) and Eq.(\ref{belFastRate}), respectively, evaluated for $n_p = n_{\mathrm{eqtr}}$ and $n_p = n_{\mathrm{pol}}$. 

Example latitudinal profiles of the electron ionization rate, normalized to 1~AU by $R^2$ and calculated for 1~AU and 5~AU, are shown in Fig.~\ref{betaElLati} for solar minimum and maximum conditions. The figure illustrates the rapid decrease in the electron ionization rate with increasing heliocentric distance and the evolution in latitudinal shape of the rate during the solar cycle.
\end{appendix}

\begin{acknowledgements}
This study is partly based on results of the International Space Science Institute Team on ``Effects of Heliospheric Breathing Due to Solar Cycle Variations from Back-Scattered Ly-Alpha'', whose support and friendliness is gratefully acknowledged. The SOLAR2000 Research Grade historical irradiances are provided courtesy of W.~Kent Tobiska and SpaceWx.com. These historical irradiances have been developed with funding from the NASA UARS, TIMED, and SOHO missions. The OMNI data were obtained from the GSFC/SPDF OMNIWeb interface at http://omniweb.gsfc.nasa.gov. This research was supported by the Polish MSRiT grants 1P03D00927 i N522 002 31/0902.
\end{acknowledgements}

\bibliographystyle{aa}
\bibliography{iplbib}
\end{document}